\documentclass[aps,preprint, amsmath,amssymb, groupedaddress]{revtex4-1}

\usepackage{graphicx}
\usepackage{dcolumn}
\usepackage{bm}
\usepackage{hyperref}
\usepackage{caption}
\usepackage{multirow}
\usepackage{subfigure}
\makeatletter
\newcommand\footnoteref[1]{\protected@xdef\@thefnmark{\ref{#1}}\@footnotemark}
\makeatother

\begin{document}


\title{Non-linear neoclassical model for poloidal asymmetries in tokamak pedestals: diamagnetic and radial effects included}

\author{Silvia Espinosa}
\email[]{sesp@mit.edu, seg@nyu.edu}
\affiliation{%
NYU Courant Institute of Mathematical Sciences, New York, NY 10012, USA
}%
\thanks{Supported by `La Caixa' and `Rafael del Pino' Fellowships, DOE Grant DE-FG0291ER54109 and NSF Grant RTG/DMS-1646339.}
\author{Peter J. Catto}
\affiliation{%
MIT Plasma Science and Fusion Center, Cambridge, MA 02139, USA
}%


\date{\today}

\begin{abstract} 
Stronger impurity density in-out poloidal asymmetries than predicted by the most comprehensive neoclassical models have been measured in several tokamaks around the world during the last decade, calling into question the reduction of turbulence by sheared radial electric  fields in H-mode tokamak pedestals. However, these pioneering theories neglect the impurity diamagnetic drift, or fail to retain it self-consistently; while recent measurements indicate that it can be of the same order as the ExB drift. We have developed the  first self-consistent theoretical model retaining the impurity diamagnetic  flow and the two-dimensional features it implies due to its associated non-negligible radial  flow divergence. It successfully explains collisionally the experimental impurity density, temperature and radial electric  field in-out asymmetries; thus making them consistent with H-mode pedestal turbulence reduction.
\end{abstract}


\maketitle

\section{Introduction and motivation}

It has been suggested~\cite{Connor2000,Burrell1997,Biglari1990,Lehnert1966,Ritz1990} that the sudden transition between states of low and high confinement, the L-H transition, involves the reduction of turbulence in the  pedestal by sheared radial electric fields. 
Indeed, this presence of strong $\mathbf{E} \times \mathbf{B}$ shear may explain the improved energy confinement observed in both H and I modes~\cite{Espinosa2018}.
For H-mode pedestals, the amount of turbulence may be only large enough to affect phenomena higher order in the gyroradius expansion, such as heat transport. Neoclassical collisional theory may then be expected to properly treat lower order phenomena, such as flows within the flux surface.

During the last two decades, the state-of-the-art neoclassical pedestal theories for collisional non-trace impurity behavior~\cite{Helander1998,Fulop1999,Fulop2001,Fulop2001b,Landreman2011, Romanelli1998, Angioni2014b, Espinosa2017, Espinosa2017b, Espinosa2018} have been analyzing the impurity parallel dynamics independently.
In other words, the physical phenomena included were selected~\cite{Helander1998} such that the interesting effects of the impurity radial flow could be self-consistently neglected for simplicity when evaluating its parallel flow and treating conservation equations.
The key simplifying assumption towards neglecting the impurity radial flow relies on taking the pedestal characteristic length to be of the same order for both impurity and main ion densities.
In this way the diamagnetic flow can be neglected for the high charge state impurity, while it is retained for the bulk ions.


These existing theories~\cite{Helander1998,Fulop1999,Fulop2001,Fulop2001b,Landreman2011, Romanelli1998, Angioni2014b, Espinosa2017, Espinosa2017b, Espinosa2018} continue to provide valuable insight into the poloidal rearrangement within a flux function of a single non-trace impurity species in thermal equilibrium with weakly poloidally varying background ions. 
For instance,
Helander~\cite{Helander1998} proved theoretically that impurities can accumulate on the inboard side of a pedestal flux surface in agreement with experimental observations~\cite{Churchill2013,Theiler2014,Churchill2015,Putterich2012,Viezzer2013,Ingesson2000}.
On the one hand, his model self-consistently assumes impurity and main ion flows significantly weaker than the impurity thermal velocity in order to neglect the impurity centrifugal force. 
If the impurity toroidal rotation was large enough, the inertial term should be retained and the centrifugal force could overtake the previous phenomena, causing the highly-charged impurities to concentrate on the low field side~\cite{Fulop1999, Angioni2014b}.
On the other hand, Helander's model~\cite{Helander1998} allows the friction of the impurities with the background ions to compete with the potential and pressure gradient terms in the impurity parallel momentum conservation equation.
The drive for the impurity density poloidal variation is given by the poloidal variation of the magnetic field in the friction term, which explains a larger impurity density on the high field side.
The original model with banana regime main ions~\cite{Helander1998} was extended to the Pfirsch-Schl\"{u}ter~\cite{Fulop2001} and plateau~\cite{Landreman2011} collisionality regimes; not only for  completeness but also in the hope of explaining larger impurity concentration on the high field side.

Charge-exchange recombination spectroscopy is used to measure the outboard (LFS) and inboard (HFS) impurity temperature, density and mean flow radial profiles in the midplane pedestal region of tokamaks such as Alcator C-Mod~\cite{McDermott2009,Churchill2013b} and ASDEX-Upgrade~\cite{Putterich2012}.
High confinement mode edge pedestals on Alcator C-Mod~\cite{Churchill2013,Theiler2014,Churchill2015} exhibit substantially stronger boron poloidal variation than predicted by the most comprehensive neoclassical theoretical models developed to date~\cite{Helander1998,Fulop2001,Fulop2001b,Landreman2011}.
Indeed, the accumulation of boron density on the high field side is up to six fold  for pressure alignment (see Fig.~1 and 6 of Ref.~\cite{Churchill2015}) and even substantially larger when taking the impurity temperature as a flux function instead.
This either calls into question the reduction of turbulence by sheared radial electric fields in H-mode tokamak pedestals or
indicates that there may be 
some physical process missing from these models.
This phenomenon may be amplifying the magnetic field in-out asymmetry, which is the only drive in previous theories, or acting as an additional drive.
Impurity peaking at the inboard side is also observed in other tokamaks, such as ASDEX-Upgrade~\cite{Putterich2012,Viezzer2013} and JET~\cite{Ingesson2000}.
In addition, up-down asymmetries have also been detected on tokamaks, such as Alcator A~\cite{Terry1977}, PLT~\cite{Burrell1979}, PDX~\cite{Brau1983}, ASDEX~\cite{Smeulders1986}, Compass-C~\cite{Durst1992} and Alcator C-Mod~\cite{Rice1997,Pedersen2002,Marr2010}. 

This pedestal impurity poloidal variation can be related to impurity accumulation. 
Helander proposed~\cite{Helander1998} that the impurities rearrange on a flux function to diminish the parallel friction with the background ions. 
By using impurity toroidal momentum, Eq. (13) in~\cite{Helander1998}, it can be shown then that this parallel friction affects the pedestal impurity radial flow.
If the total flow is inwards, highly charged divertor impurities can be absorbed through  the pedestal and accumulate in the core of the plasma.
High impurity confinement can lead to large radiative energy losses~\cite{Hender2016} that compromise the performance of high charge number metal wall tokamaks, such as Alcator C-Mod~\cite{Lipschultz2001}, ASDEX-Upgrade~\cite{Neu2009} and the JET ITER-Like Wall~\cite{Matthews2011,Neu2013,Angioni2014}.

Recently, the first method to measure the neoclassical radial impurity flux directly from available diagnostics, such as CXRS and Thomson scattering, was proposed~\cite{Espinosa2017,Espinosa2017b}.
One of its main advantages is that it bypasses the computationally demanding kinetic calculation of the full bulk ion response.
Difficult to evaluate main ion non-Maxwellian kinetic features need not be evaluated when impurity poloidal flow measurements are available.
The procedure in~\cite{Espinosa2017b} allows the inclusion of impurity seeding, ion cyclotron resonance minority heating (ICRH) and toroidal rotation effects that can be used to actively and favorably modify the radial impurity flux to prevent impurity accumulation, as explained in~\cite{Espinosa2017}.
Moreover, thanks to the measuring technique developed, the outward radial impurity flux in I-mode has been explained without invoking a (sometimes undetected) turbulent mode~\cite{Espinosa2018}.
A predictive theoretical neoclassical model for pedestal impurity flows that includes the effect of radial flows in the parallel dynamics may thus provide even more accurate insight on preventing impurity accumulation.

The impurity model in the following sections  proposes a
self-consistent two-dimensional theoretical neoclassical model for axisymmetric tokamak pedestals.
In contrast to the one-dimensional previous models~\cite{Helander1998,Fulop1999,Fulop2001,Fulop2001b,Landreman2011}, the impurity parallel dynamics is affected not only by flows contained in the flux surface but also by the impurity radial flows out of the flux surface.
The novel expressions for the impurity flow and conservation equations
may improve our ability to model pedestals and perhaps extend existing codes \cite{Landreman2012,Landreman2014} to a new dimension.
More importantly, this pedestal neoclassical model with radial flows may ultimately 
suggest how to better control or even avoid impurity accumulation in tokamaks such as JET and ASDEX-Upgrade.

Radial flow effects become important when the impurity density exhibits very strong radial gradients.
We achieve self-consistency by allowing the impurity diamagnetic drift to compete with the $\mathbf{E} \times \mathbf{B}$ drift, as supported by experimental observations~\cite{Theiler2014}.
Radial and diamagnetic flow effects substantially alter the parallel impurity flow.
The resulting modification in the impurity friction with the banana regime background ions impacts the impurity density poloidal variation,
by acting as an amplification factor on the magnetic field poloidal variation drive. 
It can lead to stronger poloidal variation that is in better agreement with the observations for physical values of the diamagnetic term.
 In addition, the poloidally-dependent component of the radial electric field can compete with its flux surface average for the first time.
 
\begin{table}[t]
 \caption{Novel physical phenomena included and poloidal asymmetries captured by the proposed neoclassical model in comparison to the state-of-the-art models developed to date.  \label{theoryt}}
\begin{ruledtabular}
\centering
 \begin{tabular}{@{}l@{}cc@{}}
\textbf{Experimental physics observed~\cite{Churchill2013,Theiler2014,Churchill2015}} & \textbf{Previous models \cite{Helander1998,Fulop1999,Fulop2001,Fulop2001b,Landreman2011}} & \textbf{This work} \\
\hline \vspace{-0.5cm}  \\
Single impurity species & $\checkmark$ & $\checkmark$ \\
Non-trace impurities & $\checkmark$ & $\checkmark$ \\
Diamagnetic flow effects & & $\checkmark$ \footnote{\label{inc}Although incorrectly claimed otherwise~\cite{Casson2015}, here it will be proven that both effects need to be included simultaneously for self-consistency.} \\
Radial flow effects & & $\checkmark$ \footnoteref{inc} \\
\end{tabular}
\end{ruledtabular}
\end{table}

The remaining sections are organized as follows.
Section~\ref{sec:orderings} is devoted to experimentally justifying the new physical phenomena included in the model and the corresponding orderings for the potential and species variables. 
The comprehensive range of collisionality for which the orderings are self-consistent, i.e. simultaneously verified, is also presented.
In Section~\ref{sec:ionkin}, the kinetic theory of the banana regime main ions is carefully analyzed when radial gradients of poloidally-varying variables are retained, in order to successfully calculate the parallel friction force between impurities and the background ions as a function of the impurity parallel flow. 
Section~\ref{sec:impflow} contains the calculation of the impurity flow with diamagnetic and radial flow effects using conservation of impurity particles and momentum. 
Special attention is drawn to the new sources of poloidal variation in order to provide insight into parallel and poloidal flow measurements.
It is shown in Sec.~\ref{sec:parallelmom} that the generalized parallel friction modifies the impurity parallel momentum conservation equation governing the impurity density poloidal rearrangement.
Finally, the results are summarized and discussed in Sec.~\ref{sec:analysisp4}.

%
%
%

\section{Self-consistent orderings of the edge pedestal theoretical model \label{sec:orderings}}

The theoretical pedestal model proposed here aims to explain the poloidal asymmetries in the impurity density, electrostatic potential, and impurity temperature by including the additional physical phenomena 
 summarized on Table~\ref{theoryt}.
This section is devoted to the development and experimental justification of the new orderings that are able to 
accomplish this task and provide additional physical insight within a self-consistent framework.
The range of applicability of this new model overlaps and extends that of previous models~\cite{Helander1998}.

\subsection{Impure tokamak pedestal} 
We consider an axisymmetric tokamak pedestal composed of Maxwell-Boltzmann banana regime electrons (subscript $e$) and bulk ions ($i$) with charge number $z_i \sim 1$.
The model includes a single highly charged impurity (z) with temperature $T_z \sim T_i \sim T_e$ and mass $M_z$ satisfying
\begin{eqnarray}
\sqrt{\frac{M_z}{M_i}} \sim  \sqrt{\frac{z_z}{z_i}} \gg 1.
\end{eqnarray}
This impurity is assumed to be collisional (Pfirsch-Schl\"{u}ter) and non-trace, so that
\begin{eqnarray}
\frac{z_z^2 n_z}{z_i^2 n_i} \sim \frac{\nu_{iz}}{\nu_{ii}} \sim \frac{\nu_{zz}}{\nu_{zi}} \sqrt{\frac{z_i}{z_z}} \sim \frac{\nu_{zz}}{\nu_{ii}} \left( \frac{z_i}{z_z} \right)^{\frac{3}{2}} \sim 1. 
\label{eq:alpha}
\end{eqnarray} 
Here $n$ is the species density and $\nu_{12}$ the collisional frequency of species $1$ with $2$.
The collisional frequencies between impurities and/or main ions are given in Appendix~\ref{collisionfreq}.

\subsection{Strong poloidal variation} 
The flux surface average of a quantity $Q$ is defined as
\begin{eqnarray}
\left< Q \right> 
= \frac{\oint \frac{Q d\theta}{\mathbf{B} \cdot \boldsymbol{\nabla} \theta}}{\oint \frac{d\theta}{\mathbf{B} \cdot \boldsymbol{\nabla} \theta}}
= \frac{\oint Q d\vartheta}{2 \pi};
\end{eqnarray}
with $\mathbf{B}$ the magnetic field, $\theta$ the poloidal angle variable and 
$d\vartheta = \frac{\left< \mathbf{B} \cdot \boldsymbol{\nabla} \theta \right>}{\mathbf{B} \cdot \boldsymbol{\nabla} \theta} d \theta$~\cite{Helander1998}.

A relationship between the poloidal derivative of the electrostatic potential, $\Phi$, and the electron and main ion densities can be obtained from their Maxwell-Boltzmann response, i.e. $n = \left< n \right> \exp \left[ - \frac{z e \left( \Phi - \left< \Phi \right> \right)}{T} \right]$, since their temperatures are taken to be lowest order flux functions.
The poloidal variation of the potential can also be related to that of the impurity density by subtracting from the quasineutrality equation its flux surface average, $n_e-\left< n_e \right> = z_i \left( n_i-\left< n_i \right> \right) + z_z \left( n_z-\left< n_z \right> \right)$, and taking the poloidal derivative to find
\begin{equation}
\frac{z_i e}{T_i} \frac{\partial \Phi}{\partial \theta} 
= \frac{z_z }{n_e \frac{T_i}{z_i T_e} + z_i n_i} \frac{\partial n_z}{\partial \theta}
=  \frac{z_i}{z_z} \frac{\frac{z_z^2 n_z}{ z_i^2 n_i}}{1 + \frac{n_e}{z_i n_i} \frac{T_i}{z_i T_e}} \frac{\partial \ln n_z}{\partial \theta}.
\label{eq:dPhidtheta}
\end{equation}

Moreover, the poloidal variation of the magnetic field, which is of the order of the inverse aspect ratio $\epsilon \ll 1$, is retained by considering it to be stronger than that of the potential and the electron and background ion densities.
Finally, the impurity density is allowed to exhibit the strongest poloidal variation, $\Delta$, that is taken to be of order $\sqrt{\epsilon}$ in order to keep nonlinear effects.

The orderings for the poloidal variation of the physical quantities under consideration can thus be conveniently summarised as 
\begin{widetext}
\begin{equation}
1
\gg
\left| \frac{\partial \ln n_z}{\partial \theta} \right|^2
\sim
\Delta^2
\sim
 \left| \frac{\partial \ln B}{\partial \theta} \right|
\sim
\epsilon
  \gg
\left| \frac{\partial \ln n_e}{\partial \theta} \right|
\sim 
 \left| \frac{\partial \ln n_i}{\partial \theta} \right|
\sim
\frac{z_i e}{T_i} \left| \frac{\partial \Phi}{\partial \theta}  \right|
\sim
\frac{z_i}{z_z}  \left| \frac{\partial \ln n_z}{\partial \theta} \right|.
\label{eq:ddthetaorderings}
\end{equation}
\end{widetext}
These orderings are in agreement with experimental observations~(see the right hand-side of Fig.~1 of Ref.~\cite{Churchill2015}) that show that the poloidal variation of the impurity density is significantly stronger than that of the magnetic field, radial electric field and impurity temperature.
The latter is assumed to satisfy 
\begin{equation*}
\frac{1}{\Delta^2} \frac{\partial \ln T_z}{\partial \theta} \ll 1.
\end{equation*}
As a result, the weak poloidal variation of the impurity temperature can be ignored in the parallel impurity momentum equation.

\subsection{Radial variation to retain diamagnetic flow effects} 

The characteristic length of the impurity density pedestal, $L_{n_z}$, satisfies $\rho_{pz} \ll L_{n_z} \ll \epsilon R \ll qR$. Here $\rho_p = \rho \frac{B}{B_p}$ is the poloidal ($p$) Larmor radius and $qR$ is the connection length, with safety factor $q$ and major radius $R$.
Consequently, the impurity and bulk ion mean flows are taken to be slower than the thermal speed of the impurities \cite{Helander1998,Fulop2001,Landreman2010}.

The radial electric field on Alcator C-Mod is determined by combining the independently-measured impurity contributions to the perpendicular impurity momentum equation~\cite{Theiler2014} to find
\begin{eqnarray}
E_r \approx \frac{1}{z_z e n_z} \frac{\partial p_z}{\partial r} + V_{zt} B_p - V_{zp} B_t ;
\label{eq:Erexp}
\end{eqnarray}
where $r$ is the radial coordinate, $p$ the pressure, and $V_t$ and $V_p$ the toroidal ($t$) and poloidal ($p$) components of the mean flow. 
This equation provides strong motivation for retaining in H mode the impurity diamagnetic effect, $\frac{1}{z_z e n_z} \frac{\partial p_z}{\partial r}$, since experimental measurements~(see Fig. 3 of Ref.~\cite{Theiler2014}) indicate that it can contribute more than 70\% of the radial electric field in~(\ref{eq:Erexp}) for Boron.
The $\mathbf{E}\times \mathbf{B}$ and impurity diamagnetic effects  are allowed to compete for the first time by ordering the impurity density radial variation to be stronger by a charge ratio than that of the potential and bulk ion density, leading to the following orderings for perpendicular gradients:
\begin{widetext}
\begin{equation}
\frac{1}{\rho_{pz}}
\gg 
\frac{\left|\boldsymbol{\nabla}_{\perp} n_z \right|}{n_z} 
\sim 
\frac{z_z}{z_i} \frac{\left|\boldsymbol{\nabla}_{\perp} \Phi  \right|}{\frac{T_i}{z_{i} e}}
\sim 
\frac{z_z}{z_i} \frac{\left|\boldsymbol{\nabla}_{\perp} n_i \right|}{n_i}
\sim
\frac{z_z}{z_i} \frac{\left|\boldsymbol{\nabla}_{\perp} T_i \right|}{T_i} 
\sim
\frac{z_z}{z_i} \frac{\left|\boldsymbol{\nabla}_{\perp} T_z \right|}{T_z}  
\gg
\frac{z_z}{z_i} \frac{\left|\boldsymbol{\nabla}_{\perp} B \right|}{B},
\label{eq:potdiamrhopi}
\end{equation}
\end{widetext}

\begin{figure}[t]
\centering
\includegraphics[width=0.1\textwidth]{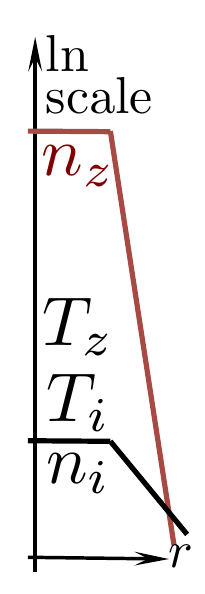}
\caption{Radial gradients schematics.}
\label{fig:raddersch}
\end{figure}
These orderings, schematized in Fig.~\ref{fig:raddersch}, are  in agreement with the experimental evidence in the right hand-side of Fig.~1 of Ref.~\cite{Churchill2015}).
Here the bulk ion temperature radial variation is taken to be as large as possible with the main ion diamagnetic effects competing but not overtaking the $\mathbf{E}\times \mathbf{B}$ contribution. 
Given that equilibration forces $T_i \sim T_z$ at all pedestal radial locations and that the radial variation of the impurity density is observed to be stronger than that of the impurity temperature~(see right hand-side on Fig.~1 of Ref.~\cite{Churchill2015}), it is reasonable to take the radial variation of both temperatures to be of the same order.

The experimental evidence of the importance of the diamagnetic effects is supported by theoretical evidence as well. 
By taking the radial derivative of the Maxwell-Boltzmann bulk ion response and of the poloidally varying piece of quasineutrality,
$z_z \left( n_z - \left< n_z \right> \right) = \left( n_e - \left< n_e \right> \right) - z_i \left( n_i - \left< n_i \right> \right)$,
a relationship between the radial variation of the poloidal part of the potential and the impurity and bulk ion densities consistent with (\ref{eq:potdiamrhopi}) can be obtained:
\begin{widetext}
\begin{eqnarray}
\frac{\frac{z_z n_z}{z_i n_i}}{1 + \frac{n_e}{z_i n_i} \frac{T_i}{z_i T_e} } \frac{1}{n_z} \frac{\partial \left( n_z - \left< n_z \right> \right)}{\partial \psi}
=
\frac{z_i e}{T_i} \frac{\partial \left( \Phi - \left< \Phi \right> \right)}{\partial \psi}
=
- \frac{1}{n_i} \frac{\partial \left( n_i - \left< n_i \right> \right)}{\partial \psi}.
\label{eq:dnztdpsipot}
\end{eqnarray}
\end{widetext}
In summary, even though the poloidally dependent components of the potential and electron and bulk ion densities are much smaller than their corresponding flux surface averages (\ref{eq:ddthetaorderings}), 
unlike in~\cite{Helander1998}, the radial derivatives of these components are allowed to compete since
\begin{widetext}
\begin{equation}
\begin{aligned}
\frac{z_i}{z_z} \frac{\left|\boldsymbol{\nabla}_{\perp} \left( n_z - \left< n_z \right> \right) \right|}{n_z} 
&\sim
\frac{\left|\boldsymbol{\nabla}_{\perp} \left( \Phi - \left< \Phi  \right> \right) \right|}{\frac{T_i}{z_{i} e}} 
\sim
\frac{\left|\boldsymbol{\nabla}_{\perp} \left( n_i - \left< n_i \right> \right) \right|}{n_i}
\sim
\frac{\left|\boldsymbol{\nabla}_{\perp} \left( n_e - \left< n_e \right> \right) \right|}{n_i}
\\
&
\sim
\Delta \frac{z_i}{z_z} \frac{\left|\boldsymbol{\nabla}_{\perp} \left< n_z \right> \right|}{n_z} 
\sim
\Delta \frac{\left|\boldsymbol{\nabla}_{\perp} \left< \Phi  \right>  \right|}{\frac{T_i}{z_{i} e}} 
\sim
\Delta \frac{\left|\boldsymbol{\nabla}_{\perp} \left<  n_i  \right> \right|}{n_i} 
\sim
\Delta \frac{\left|\boldsymbol{\nabla}_{\perp}  \left< n_e \right> \right|}{n_e}.
\end{aligned}
\label{eq:ddpsiorderings}
\end{equation}
\end{widetext}
Since the negative slope of the impurity density is more negative on the inboard side (see right hand side of Fig.~1 of Ref.~\cite{Churchill2015}), the model
predicts via (\ref{eq:ddpsiorderings}) the radial electric field be less negative on the inboard than on the outboard side.
This is consistent with the experimental observation shown in
Fig. 5 in~\cite{Theiler2014}.

\subsection{Species collisionality} 

The assumptions of having lowest-order Maxwellian impurities, (\ref{eq:f1gacollfreq}); bulk and impurity temperature equilibration, (\ref{eq:compheatvseq}); banana regime, (\ref{eq:constraint3banana}), Maxwell-Boltzmann bulk ions, (\ref{eq:ionMBconst}); and friction not affecting the lowest-order perpendicular impurity flow, (\ref{eq:perpmomfricrad}), are checked a posteriori. 
These are the most restrictive inequalities and are obtained latter for the equation numbers given above. 
Doing so leads to the conclusion that self-consistency is satisfied when
\begin{widetext}
\begin{eqnarray}
\min \left\{ 1, \sqrt{\frac{z_i}{z_z}} \frac{L_{n_z}}{\rho_{pz}} \right\}
\gg
\Delta \frac{\lambda_z}{qR}
\gg
\max \left\{
\frac{\Delta}{{\epsilon}^{\frac{3}{2}} } \frac{z_i^2}{z_z^2},
 \frac{\rho_{pz}}{L_{n_z}} \sqrt{\frac{z_z}{z_i}} \frac{z_i^2}{z_z^2},
 \frac{\rho_{pz}}{L_{n_z}} \sqrt{\frac{z_z}{z_i}} \frac{B_p^2}{B^2}
\right\}.
\label{eq:collisionalityrange}
\end{eqnarray}
\end{widetext}
Here $\lambda$ is the mean free path, which is the thermal speed $v_{T}=\sqrt{\frac{2 T}{M}}$ divided by the sum of the like and unlike collision frequencies. 
Note also that the ratio of impurity to background ion mean free paths is given by
\begin{eqnarray}
\frac{\lambda_z}{\lambda_i}
\sim
\frac{\frac{v_{Tz}}{\nu_{zz}+ \nu_{zi}}}{\frac{v_{Ti}}{\nu_{ii}+ \nu_{iz}}}
\sim
\left( \frac{z_i}{z_z} \right)^2,
\label{eq:lambdazlambdairatio}
\end{eqnarray}
where $\frac{z_z^2 n_z}{z_i^2 n_i} \sim 1$ 
 has been used to relate the collisional frequencies.
In the collisional range under consideration (\ref{eq:collisionalityrange}), the  self-collisional frequencies are much smaller than the gyrofrequency $\Omega = \frac{z e B}{cM} = \frac{v_T}{\rho}$, which is of the same order for main and impurity ions.

The assumptions of having lowest-order Maxwellian impurities and friction not affecting the lowest-order perpendicular impurity flow
 are checked in Appendices \ref{sec:paper4_a3} and \ref{sec:checking_a}, respectively. 
The bulk and impurity temperature equilibration is checked in Appendix \ref{sec:temperatureeq}.

The condition to have banana regime background ions,
\begin{equation*}
 \frac{qR}{\epsilon^{\frac{3}{2}} \lambda_i}
\ll 1, 
\end{equation*}
could be rewritten by using (\ref{eq:lambdazlambdairatio}) as a function of the impurity mean free path as follows:
\begin{equation}
 \frac{\lambda_z}{qR} \epsilon^{\frac{3}{2}} \frac{z_z^2}{z_i^2} 
\gg 1.
\label{eq:constraint3banana}
\end{equation}


The Maxwell-Boltzmann behaviour for the main ions is obtained if $T_i$ is a flux function and the bulk ion pressure and potential gradients are the dominant terms in the parallel momentum equation for the background ions,
which requires
\begin{eqnarray}
\frac{R_{iz\parallel}}{\nabla_{\parallel} p_i + z_i e n_i \nabla_{\parallel} \Phi}
 \sim
\frac{
\nu_{iz} n_i M_i v_{Tz} \frac{\rho_{pz}}{L_{n_z}} 
}{
\Delta \frac{z_i}{z_z}  \frac{p_i}{qR}  
}
\sim 
\frac{1}{\Delta}  \frac{qR}{\lambda_z}  \frac{\rho_{pz}}{L_{n_z}} \left( \frac{z_i}{z_z} \right)^{\frac{3}{2}}
\ll 1.
\label{eq:ionMBconst}
\end{eqnarray}

\section{First-order main ion kinetic equation and friction force \label{sec:ionkin}}

The gyroaveraged background ion distribution function is given by $\bar{f}_i = f_{iM\left<\right>}  +  \bar{f}_{i1}$ \cite{Helander1998, Landreman2011}.
The lowest order distribution function can be chosen to be a stationary Maxwellian and a flux function:
\begin{eqnarray}
 f_{iM\left<\right>} 
=  \left< n_i \right> \left( \psi \right) \left[ \frac{M_i}{2 \pi T_i \left( \psi \right)} \right]^{\frac{3}{2}} 
\exp \left[ - \frac{M_i v_i^2}{2T_i \left( \psi \right)} \right].
\label{eq:fMfsa}
\end{eqnarray}
The gyrophase independent first-order correction is proven~\cite{Helander2005} to be given by
\begin{eqnarray}
v_{i\parallel} \nabla_{\parallel}
\left(  
\bar{ f}_{i1}
+ \frac{I v_{i\parallel}}{\Omega_i} \left. \frac{\partial f_{iM\left<\right>}}{\partial \psi} \right|_{E_{i\left<\right>}}
\right)
+ \frac{z_i e v_{i\parallel} }{T_i} f_{iM\left<\right>} \nabla_{\parallel} \Phi 
=
C_{ii1} \left\{ \bar{f}_{i1} \right\} 
+ C_{iz1} \left\{ \bar{f}_{i1} \right\}.
\end{eqnarray}
Here the spatial gradients are taken keeping constant the magnetic moment, $\mu_i=\frac{v_{i\perp}^2}{2B}$, and the lowest-order total energy, 
$
E_{i\left<\right>} = \frac{v_i^2}{2} + \frac{z_i e \left< \Phi \right>}{M_i}
\label{eq:Eifsa}
$.
In addition, the linearized gyroaveraged unlike collision operator of bulk ions with lowest-order drifting Maxwellian impurities is
\begin{eqnarray}
C_{iz1} \left\{ \bar{f}_{i1} \right\}
= \frac{3 \sqrt{2 \pi} \nu_{iz} T_i^{\frac{3}{2}}}{4 M_i^{\frac{3}{2}}}
\left[
\boldsymbol{\nabla}_{v_i} \cdot 
\left(
\boldsymbol{\nabla}_{v_i}  \boldsymbol{\nabla}_{v_i}  v_i
\cdot
\boldsymbol{\nabla}_{v_i} \bar{f}_{i1} 
\right)
+ 2  f_{iM\left<\right>} \frac{M_i}{T_i} \frac{v_{i\parallel}}{v_i^3} V_{z\parallel}  
\right].
\label{eq:Ciz1fsa}
\end{eqnarray}

The parallel friction force between impurities and the background ions is calculated~\cite{Helander2005} by taking the parallel first order moment of the unlike collision operator (\ref{eq:Ciz1fsa}) to be given by
 \begin{equation}
 \begin{aligned}
R_{zi\parallel}
= 
 -  \frac{\left< p_i \right> I \nu_{iz}}{\Omega_i}
  \left[  
  \frac{d \ln  \left< p_i \right>}{d \psi} 
+  \frac{z_i e}{T_i} \frac{d \left<\Phi \right>}{d \psi}
-\frac{3}{2} \frac{ d \ln T_i }{d \psi}
 \right]
+M_i \nu_{iz} \left< n_i \right> 
\left( \frac{n_i}{\left< n_i \right>} \frac{u B}{n_i} - V_{z\parallel} \right),
\end{aligned}
\end{equation}
where for general collisionality main ions
\begin{eqnarray}
u 
= 
\frac{3 \sqrt{\pi}}{\sqrt{2}} \frac{T_i^{\frac{3}{2}}}{ M_i^{\frac{3}{2}}}
\int \frac{h_i}{B} \frac{v_{i\parallel}}{v_i^3} d^3 v_i
\end{eqnarray}
with
\begin{eqnarray}
h_i
=
\bar{ f}_{i1}
+ \frac{I v_{i\parallel}}{\Omega_i} \left. \frac{\partial f_{iM\left<\right>}}{\partial \psi} \right|_{E_{i\left<\right>}}
+ \frac{z_i e \left( \Phi - \left< \Phi \right> \right)}{T_i} f_{iM\left<\right>}.
\label{eq:hi}
\end{eqnarray}
In particular, for banana bulk ions $h_i$ does not depend on poloidal angle but via $v_{i\parallel}/\left| v_{i\parallel} \right|$.
Consequently, $u$ is a flux function since $d^3 v_i \propto \frac{B}{v_{i\parallel}} d \mu_i d E_{i\left<\right>}$ \cite{Helander1998}.


\section{Impurity flow \label{sec:impflow} }

This section is devoted to the calculation of the pedestal impurity flux including diamagnetic and non-diffusive radial flow effects self-consistently.
To begin with, the perpendicular impurity flow is obtained from perpendicular momentum conservation for the impurities.
Next, the impurity continuity equation is solved for the form of the parallel impurity flow within a flux function. 
The divergence of the radial flow has to be cleverly rearranged to facilitate the integration of the continuity equation.  
Finally, the parallel momentum equation for the impurities is considered and its solubility condition used to determine the unknown flux function. 
The friction with the background ions is modified due to 
retention of both the impurity diamagnetic flow and the radial flow effects that modify the
the impurity parallel mean flow.

\subsection{Perpendicular impurity flow \label{subsec:perpimpflow}}

The momentum conservation for impurities balances electrostatic, magnetic, isotropic and anisotropic pressure forces, inertia and friction with the background:
\begin{widetext}
\begin{eqnarray}
z_z e n_z \left( \boldsymbol{\nabla} \Phi - \frac{\mathbf{V}_z \times \mathbf{B}}{c} \right) 
+ \boldsymbol{\nabla} p_z 
+ \boldsymbol{\nabla} \cdot \boldsymbol{\pi}_z 
+ M_z n_z \mathbf{V}_z \cdot \boldsymbol{\nabla} \mathbf{V}_z 
= \mathbf{R}_{zi} .
\label{eq:momconsvimp}
\end{eqnarray}
\end{widetext}
It is reasonable to assume that the perpendicular velocity is dominated by the $\mathbf{E}\times \mathbf{B}$, as in~\cite{Helander1998}, and the new diamagnetic drift, since they are allowed to compete.
The orderings (\ref{eq:ddthetaorderings}), (\ref{eq:potdiamrhopi}), (\ref{eq:ddpsiorderings}) and (\ref{eq:collisionalityrange}) are chosen to make the perpendicular projection of the inertia, friction and divergence of the anisotropic pressure tensor negligible; as can be checked a posteriori in Appendix \ref{sec:checking_a}.
Therefore,
\begin{widetext}
\begin{equation}
\begin{aligned}
\mathbf{V}_{z\perp} 
=
\frac{c}{B^2} \mathbf{B} \times \left( \boldsymbol{\nabla} \Phi + \frac{\boldsymbol{\nabla} p_z}{z_z e n_z} \right) 
= &
 \frac{c}{B^2} \mathbf{B} \times \boldsymbol{\nabla} \psi \left( \frac{\partial \Phi}{\partial \psi} + \frac{1}{z_z e n_z} \frac{\partial p_z}{\partial \psi} \right) 
 \\
&+ \frac{c}{B^2} \mathbf{B} \times \boldsymbol{\nabla} \theta
 \left( \frac{\partial \Phi}{\partial \theta} + \frac{1}{z_z e n_z} \frac{\partial p_z}{\partial \theta} \right) .
 \end{aligned}
 \label{eq:Vzperp}
\end{equation}
\end{widetext}
The axisymmetric tokamak magnetic field is taken to be
\begin{eqnarray}
\mathbf{B} = I \boldsymbol{\nabla} \zeta + \boldsymbol{\nabla} \zeta \times \boldsymbol{\nabla} \psi;
\label{eq:Bfield}
\end{eqnarray}
where $\zeta$ is the toroidal angle,
 $2 \pi \psi$ the poloidal magnetic flux with $\left| \boldsymbol{\nabla} \psi \right| = R B_p$
and $I \left( \psi \right) = RB_t$ a flux function.
From (\ref{eq:Bfield}), it follows that
  \begin{eqnarray}
\mathbf{B} \times \boldsymbol{\nabla} \psi 
= I \mathbf{B} - B^2 R^2 \boldsymbol{\nabla} \zeta
\label{eq:Bxpsi}
\end{eqnarray}
and
  \begin{eqnarray}
\mathbf{B} \times \boldsymbol{\nabla} \theta \cdot \boldsymbol{\nabla} \psi 
= - I \mathbf{B} \cdot \boldsymbol{\nabla} \theta .
\label{eq:Bxthetapsi}
\end{eqnarray}

The projections of the perpendicular impurity mean flow (\ref{eq:Vzperp}) in the directions perpendicular to the flux surface (referred to as \emph{radial})
and within the surface but perpendicular to the magnetic field are evaluated (\ref{eq:Bxthetapsi}) to respectively find
\begin{equation}
\mathbf{V}_{z\perp} \cdot \frac{\boldsymbol{\nabla} \psi}{R B_p}
= 
- \frac{cI}{B^2} \frac{\mathbf{B} \cdot \boldsymbol{\nabla} \theta}{R B_p}
 \left( \frac{\partial \Phi}{\partial \theta} + \frac{1}{z_z e n_z} \frac{\partial p_z}{\partial \theta} \right) 
 \sim
\Delta \frac{\rho_{pz}}{q R} v_{Tz} 
\label{eq:Vzperprad}
\end{equation}
 and
 \begin{widetext}
 \begin{equation}
 \begin{aligned}
 \mathbf{V}_{z\perp} \cdot \frac{\mathbf{B} \times \boldsymbol{\nabla} \psi }{B R B_p}
= &
 c \frac{R B_p}{B} \left( \frac{\partial \Phi}{\partial \psi} + \frac{1}{z_z e n_z} \frac{\partial p_z}{\partial \psi} \right) 
&&  \sim
\frac{\rho_{z}}{L_{n_z}} v_{Tz}
 \\
&+ c \frac{\boldsymbol{\nabla} \psi \cdot \boldsymbol{\nabla} \theta}{B R B_p}
 \left( \frac{\partial \Phi}{\partial \theta} + \frac{1}{z_z e n_z} \frac{\partial p_z}{\partial \theta} \right) 
&& \sim
s \Delta \frac{\rho_{pz}}{qR} v_{Tz} .
\end{aligned}
\label{eq:Vzperppol}
\end{equation}
\end{widetext}
The estimated size of the terms is shown to the right of the terms in (\ref{eq:Vzperprad}) and (\ref{eq:Vzperppol}), where $s$ is the magnetic flux surface shape factor:
\begin{eqnarray}
0 \lesssim s = r \boldsymbol{\nabla} \theta \cdot \frac{\boldsymbol{\nabla} \psi}{ R B_p}   \lesssim 1.
\end{eqnarray} 
Even thought this factor is small in the concentric circle flux surface large aspect ratio limit, it is retained in this calculation for further accuracy.

\subsection{Parallel impurity flow} 

\begin{figure}[t]
\centering
\subfigure[\label{fig:flowImp1Dr_new} Parallel dynamics can be individually analyzed in a one dimensional model without diamagnetic effects.]{
\includegraphics[width=0.45\textwidth]{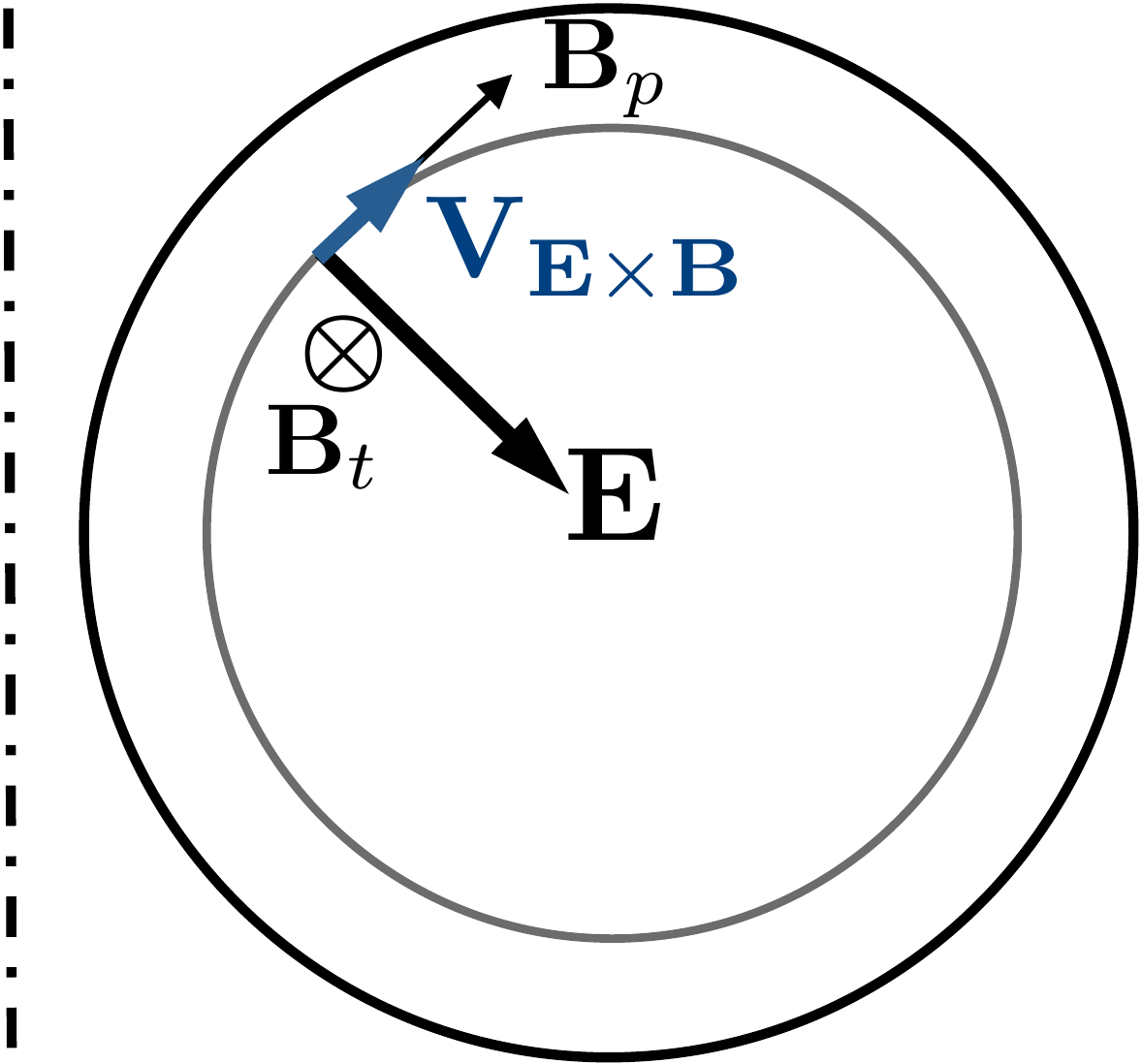}
}
\hfill
\subfigure[\label{fig:flowImp2Dr_new} Two dimensional effects given by the radial flow must be retained due to its large divergence to include impurity diamagnetic effects.
Diamagnetic effects are included in the electric field, $\mathbf{E^*_{\mbox{new}}} = - \left( \boldsymbol{\nabla} \Phi + \frac{\boldsymbol{\nabla} p_z}{z_z e n_z} \right)$.]{
\includegraphics[width=0.45\textwidth]{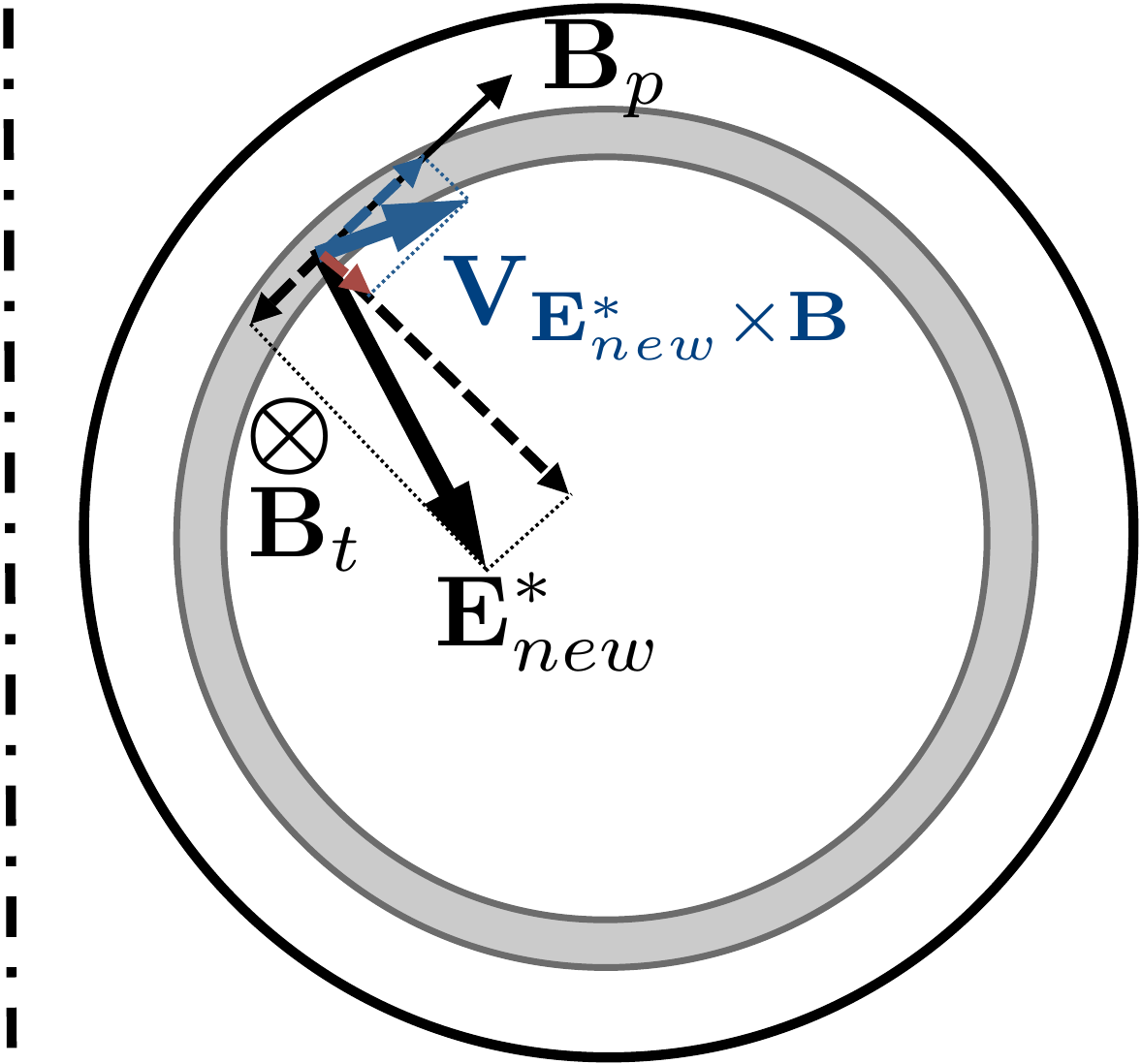}
}
\caption{Schematic of the perpendicular impurity flow, given by the $\mathbf{E} \times \mathbf{B}$ drift, that affects the parallel impurity flow in an tokamak cross section.}
\end{figure}

The relationship between the perpendicular and parallel impurity flows must satisfy the conservation of particles equation,
\begin{eqnarray}
\boldsymbol{\nabla} \cdot \left( n_z \mathbf{V}_z \right)
= \boldsymbol{\nabla} \cdot \left( n_z \mathbf{V}_{z\perp} \right) + \mathbf{B} \cdot \boldsymbol{\nabla} \left( V_{z\parallel} \frac{n_z}{B} \right) = 0.
\label{eq:conspartparperp}
\end{eqnarray}
The divergence of the perpendicular flux in an axisymmetric tokamak is given by
\begin{widetext}
\begin{eqnarray}
\boldsymbol{\nabla} \cdot \left( n_z \mathbf{V}_{z\perp} \right)
= 
\mathbf{B} \cdot \boldsymbol{\nabla} \theta
\left[ 
 \frac{\partial}{\partial \psi} \left( \frac{n_z \mathbf{V}_{z\perp} \cdot \boldsymbol{\nabla} \psi}{\mathbf{B} \cdot \boldsymbol{\nabla} \theta} \right)
+ 
 \frac{\partial}{\partial \theta} \left( \frac{n_z \mathbf{V}_{z\perp} \cdot \boldsymbol{\nabla} \theta}{\mathbf{B} \cdot \boldsymbol{\nabla} \theta} \right)
\right] .
\label{eq:divperpflux}
\end{eqnarray}
\end{widetext}
The two components of the impurity flow, as given by (\ref{eq:Vzperprad}) and (\ref{eq:Vzperp}), result in comparable contributions to the divergence when the impurity diamagnetic flow is retained as can be seen from
\begin{widetext}
\begin{equation}
\frac{\partial}{\partial \psi} \left(
\frac{ n_z \mathbf{V}_{z\perp} \cdot \boldsymbol{\nabla} \psi}{\mathbf{B} \cdot \boldsymbol{\nabla} \theta}
\right)
\sim
\frac{\partial}{\partial \theta} \left(
\frac{ n_z \mathbf{V}_{z\perp} \cdot \boldsymbol{\nabla} \theta }{\mathbf{B} \cdot \boldsymbol{\nabla} \theta}
\right)
 = 
 \frac{\partial}{\partial \theta} \left[
  \frac{cI}{z_z e B^2} \left( z_z e n_z \frac{\partial \Phi}{\partial \psi} + \frac{\partial p_z}{\partial \psi} \right)
  \right] .
  \label{eq:divVzperpord}
\end{equation}
\end{widetext}
The preceding implies that the parallel dynamics depends on the perpendicular dynamics (Fig.~\ref{fig:flowImp2Dr_new}), in contrast to all the previous models~\cite{Helander1998} (Fig.~\ref{fig:flowImp1Dr_new}). 
In other words, the impurity radial flow affects the parallel flow when the diamagnetic flow is retained.

The physical phenomena included in the model, (\ref{eq:dPhidtheta}) and (\ref{eq:ddthetaorderings}), have been purposely selected in order to make feasible the integration of the conservation of particles equation (\ref{eq:conspartparperp}) to determine the impurity parallel flow.
The first step towards expressing the divergence of the radial impurity flux (\ref{eq:Vzperprad})
in the form of a parallel gradient of a scalar consists of using the relationships between the poloidal variation of the potential and impurity density in~(\ref{eq:dPhidtheta}) to find
\begin{eqnarray}
\frac{\partial p_z}{\partial \theta}
+
z_z e n_z \frac{\partial \Phi}{\partial \theta}
=
\frac{\partial P}{\partial \theta}
+ O \left( \Delta^2 \frac{z_i}{z_z} p_z \right);
\label{eq:dPdtheta} 
\end{eqnarray}
where
 \begin{eqnarray}
P
= \left( p_z - \left< p_z \right> \right) 
 + \frac{z_z^2 T_i}{z_i^2 n_i \left( 1 + \frac{n_e}{z_i n_i} \frac{T_i}{z_i T_e} \right)} 
\left[ 
\left< n_z \right> \left( n_z - \left< n_z \right> \right) 
+ \frac{\left( n_z - \left< n_z \right> \right)^2 }{2}
\right] .
\label{eq:P} 
\end{eqnarray}
Second, even though both impurity density and magnetic field poloidal variations are retained (\ref{eq:ddthetaorderings}), their product $\epsilon \Delta$ is assumed to be negligible,
primarily to bring the magnetic field under the poloidal derivative to lowest order by writing
\begin{equation}
\frac{I}{B^2} \frac{\partial P}{\partial \theta} 
=
\frac{\partial}{\partial \theta} \left( \frac{I P}{\left< B^2 \right>} \right)
 +
  \frac{I}{B^2} \left( 1 - \frac{B^2}{\left< B^2 \right>} \right) \frac{\partial P}{\partial \theta}
 =
\frac{\partial}{\partial \theta} \left( \frac{I P}{\left< B^2 \right>} \right)
 + 
 O \left(  \epsilon \Delta \frac{I p_z}{B^2} \right).
 \label{eq:Binsidepolder}
 \end{equation}

By using (\ref{eq:Vzperprad}), (\ref{eq:divVzperpord}), (\ref{eq:dPdtheta}) and (\ref{eq:Binsidepolder}), the lowest order conservation of impurity particles equation (\ref{eq:conspartparperp}) to order $\epsilon$  can hence be rewritten as
 \begin{eqnarray}
\mathbf{B} \cdot \boldsymbol{\nabla} 
\left[
\frac{n_z V_{z\parallel}}{B}
+ \frac{c I}{z_z e B^2} \left( \frac{\partial p_z}{\partial \psi} + z_z e n_z \frac{\partial \Phi}{\partial \psi} \right)
-   \frac{c }{z_z e} \frac{\partial}{\partial \psi} \left( \frac{I P}{\left< B^2 \right>} \right)
\right]
= 0.
\label{eq:conspartfinal}
 \end{eqnarray}
The parallel impurity flow is then obtained by integrating in poloidal angle to find
\begin{eqnarray}
V_{z\parallel}
=
\frac{B K_z}{n_z} 
- \frac{c I}{z_z e B n_z} \left( \frac{\partial p_z}{\partial \psi} + z_z e n_z \frac{\partial \Phi}{\partial \psi} \right)
+  \frac{c B}{z_z e n_z} \frac{\partial}{\partial \psi} \left( \frac{I P}{\left< B^2 \right>} \right)
\sim
\frac{\rho_{pz}}{L_{n_z}} v_{Tz},
\label{eq:Vz||ord}
\end{eqnarray}
where $K_z \left( \psi \right)$ is an unknown flux function. 
In conclusion, the impurity flow is given by
\begin{equation}
\begin{aligned}   
\mathbf{V}_z
 = \mathbf{V}_{z \perp} + V_{z \parallel} \frac{\mathbf{B}}{B} 
 = &
  \frac{\mathbf{B} }{n_z}
  \left[
   K_z
+  \frac{c }{z_z e} \frac{\partial}{\partial \psi} \left( \frac{I P}{\left< B^2 \right>} \right)
  \right]
  - c R^2  \boldsymbol{\nabla} \zeta 
  \left(
\frac{\partial \Phi}{\partial \psi} + \frac{1}{z_z e n_z} \frac{\partial p_z}{\partial \psi}
\right)
 \\ 
 &  
   +
   \frac{cI }{B^2} 
\boldsymbol{\nabla} \zeta \times \boldsymbol{\nabla} \theta  
 \left( \frac{\partial \Phi}{\partial \theta} + \frac{1}{z_z e n_z} \frac{\partial p_z}{\partial \theta} \right) ,
 \end{aligned}
 \label{eq:Vzfinal}
\end{equation}
where (\ref{eq:Bfield}) and (\ref{eq:Bxpsi}) have been used to evaluate the perpendicular impurity flow (\ref{eq:Vzperp}).

\subsection{Calculation of the integration constant in the parallel impurity flow}

Let us now turn our attention to the parallel impurity momentum conservation equation,
\begin{eqnarray}
{\nabla}_{\parallel} p_z + z_z e n_z {\nabla}_{\parallel} \Phi 
+ M_z n_z \mathbf{V}_z \cdot \boldsymbol{\nabla} \mathbf{V}_z \cdot \frac{\mathbf{B}}{B} 
+ \boldsymbol{\nabla} \cdot \boldsymbol{\pi}_z \cdot \frac{\mathbf{B}}{B} 
= R_{zi \parallel},
\end{eqnarray}
which is dominated by the friction and the impurity pressure and potential gradients.
The inertial and diagonal (subscript $d$) and off-diagonal (subscript $g$) viscous forces can be neglected according to our orderings, since
\begin{equation}
\frac{
M_z n_z \mathbf{V}_z \cdot \boldsymbol{\nabla} \mathbf{V}_z \cdot \frac{\mathbf{B}}{B} 
}{
{\nabla}_{\parallel} p_z + z_z e n_z {\nabla}_{\parallel} \Phi
} 
\sim
\frac{
M_z n_z \mathbf{V}_z \cdot \boldsymbol{\nabla} V_{z\parallel}
}{
\frac{\mathbf{B}}{B} \cdot \boldsymbol{\nabla} \theta \left( \frac{\partial p_z}{\partial \theta} + z_z e n_z \frac{\partial \Phi}{\partial \theta} \right) 
} 
\sim
\frac{
M_z n_z \Delta  \frac{v^2_{Tz}}{qR} \frac{{\rho}_{pz}^2}{L_{n_z}^2} 
}{\Delta \frac{p_z}{qR}} 
\sim
\frac{{\rho}_{pz}^2}{L_{n_z}^2} 
\ll 1.
\label{eq:neglinertia||mom}
\end{equation}
\begin{equation}
\frac{
\boldsymbol{\nabla} \cdot \boldsymbol{\pi}_{zdC} \cdot \frac{\mathbf{B}}{B}
}{
{\nabla}_{\parallel} p_z + z_z e n_z {\nabla}_{\parallel} \Phi
} 
\sim
\max \left\{
\Delta  \frac{\rho_{pz}}{L_{n_z}}  \frac{\lambda_z}{ qR},
\Delta \frac{\lambda_z^2}{q^2R^2} \frac{\partial \ln T_z}{\partial \theta} ,
\left( \frac{\rho_{pz}}{qR}  \frac{\partial \ln T_z}{\partial \theta}  \right)^2
\right\}
 \ll 1
 \label{eq:neglviscd||mom}
\end{equation}
and
\begin{equation}
\frac{
\boldsymbol{\nabla} \cdot \boldsymbol{\pi}_{zgC} \cdot \frac{\mathbf{B}}{B}
}{
{\nabla}_{\parallel} p_z + z_z e n_z {\nabla}_{\parallel} \Phi
} 
\sim
 \max \left\{
\frac{\rho^2_{pz}}{L^2_{n_z}}, 
\frac{\rho_{pz} }{L_{n_z}} \frac{\lambda_z}{qR}  \frac{\partial \ln T_z}{\partial \theta}
 \right\}
 \ll 1.
   \label{eq:neglviscod||mom}
\end{equation}
Here the diagonal ($d$) and off-diagonal or gyroviscous ($g$) part of the viscous tensor are obtained on Eq. (42) and Eq. (44) of \cite{Catto2004} (subscript $C$), respectively. 
The precise expressions can be found in Appendix \ref{sec:viscoustensorapp}.

The calculation of the parallel friction force between impurities and the background ions, as outlined in Appendix~\ref{sec:ionkin}, leads to
 \begin{equation}
R_{zi\parallel}
=
 -  \frac{\left< p_i \right> I \nu_{iz}}{\Omega_i}
  \left[  
  \frac{d \ln  \left< p_i \right>}{d \psi} 
+  \frac{z_i e}{T_i} \frac{d \left<\Phi \right>}{d \psi}
-\frac{3}{2} \frac{ d \ln T_i }{d \psi}
 \right]
+M_i \nu_{iz} \left< n_i \right> 
\left( \frac{n_i}{\left< n_i \right>} \frac{u B}{n_i} - V_{z\parallel} \right).
 \label{eq:Rzi||Vz||}
\end{equation}
For banana main ions,
\begin{eqnarray}
u 
= 
\frac{3 \sqrt{\pi}}{\sqrt{2}} \frac{T_i^{\frac{3}{2}}}{ M_i^{\frac{3}{2}}}
\int  \frac{d^3 v_i}{B} \frac{v_{i\parallel}}{v_i^3} 
h_i \left( \psi, \frac{v_{i\parallel}}{\left| v_{i\parallel} \right|}, \mu_i =  \frac{v_{i\perp}^2}{2B}, E_{i\left<\right>} = \frac{v_i^2}{2} + \frac{z_i e \left< \Phi \right>}{M_i} \right)
\end{eqnarray}
is a flux function since $d^3 v_i \propto \frac{B}{v_{i\parallel}} d \mu_i d E_{i\left<\right>}$~\cite{Helander1998}.
Here $h_i = f_i - f_{iM}^{*} \left( \psi^{*}_i , E_i \right)$ vanishes in the trapped domain,
 where $f_i$ is the main ion distribution function. 
 The distribution $f_{iM}^{*}$ is a modification of a Maxwell-Boltzmann distribution depending only on the constants of the motion canonical angular momentum $ \psi_i^{*} = \psi - \frac{c M_i}{z_i e} R^2 \boldsymbol{\nabla} \zeta \cdot \mathbf{v}_i$, which replaces the $\psi_i$ dependence,
 and total energy $E_i = \frac{v_i^2}{2} + \frac{z_i e \Phi}{M_i}$.

The dominant terms in parallel momentum conservation, ${\nabla}_{\parallel} p_z + z_z e n_z {\nabla}_{\parallel} \Phi 
= R_{zi \parallel}$, multiplied by the magnitude of the magnetic field can be evaluated by using 
(\ref{eq:dPdtheta}) in the left hand side and (\ref{eq:Vz||ord}) and  (\ref{eq:Rzi||Vz||}) in the right hand side
to find
\begin{equation}
\begin{aligned}
  \mathbf{B} \cdot \boldsymbol{\nabla} P
 = 
 - M_i  \left< \frac{\nu_{iz}}{n_z} \right>  \left< n_i \right> 
 &
 \left[
 \frac{c I}{z_i e}  n_z T_i 
  \left(  
  \frac{d \ln  \left< p_i \right>}{d \psi} 
-\frac{3}{2} \frac{ d \ln T_i }{d \psi}
 \right)
- u B^2 \frac{n_z}{\left< n_i \right>}
+ B^2 K_z 
\right.
 \\
& 
\left.
- \frac{c I}{z_z e}  \frac{\partial p_z}{\partial \psi} 
- c I n_z \frac{\partial \left(\Phi - \left< \Phi \right> \right)}{\partial \psi}
+  \frac{c B^2}{z_z e} \frac{\partial}{\partial \psi} \left( \frac{I P}{\left< B^2 \right>} \right)
\right],
\end{aligned}
\label{eq:||momKz}
\end{equation}
where the Coulomb logarithm has been taken to be a flux function.

The unknown flux function $K_z$ in the parallel impurity flow (\ref{eq:Vz||ord}) is determined 
by flux surface averaging the parallel momentum conservation (\ref{eq:||momKz}), to find eventually that
\begin{equation}
K_z 
=
\frac{1}{ \left< B^2 \right>}
\left[
 \frac{c I}{z_z e}  \frac{\partial \left< p_z \right>}{\partial \psi} 
 -  \frac{c I}{z_i e}  \left< n_z \right> T_i 
  \left(  
  \frac{d \ln  \left< p_i \right>}{d \psi} 
-\frac{3}{2} \frac{ d \ln T_i }{d \psi}
 \right)
 + \frac{u}{\left< n_i \right>} \left< B^2 n_z \right> 
 \right].
 \label{eq:Kzbefsimpl}
\end{equation}
This form arises from the cancellation of terms justified as follows.
The following term has been neglected in (\ref{eq:Kzbefsimpl}):
\begin{eqnarray}
\frac{
- \frac{c}{z_z e} \left<  \left( B^2 - \left<  B^2 \right> \right) \frac{\partial}{\partial \psi} \left[ \frac{I \left( P - \left< P  \right> \right)}{\left< B^2 \right>} \right] \right>
}{
\frac{c I}{z_z e}  \frac{\partial \left< p_z \right>}{\partial \psi} 
}
\sim
\epsilon \Delta
\ll 1,
\end{eqnarray}
consistent with the orderings in (\ref{eq:Binsidepolder}).
Also the non-linear terms cancel each other out to lowest order in (\ref{eq:Kzbefsimpl}), 
\begin{equation}
\frac{\partial \left< P \right>}{\partial \psi} 
= 
\left<
\frac{z_z^2 T_i \left( n_z - \left< n_z \right> \right) }{z_i^2 n_i \left( 1 + \frac{n_e}{z_i n_i} \frac{T_i}{z_i T_e} \right)} 
\frac{\partial \left( n_z - \left< n_z \right> \right)}{\partial \psi}
\right>
= 
z_z e
\left<
\left( n_z - \left< n_z \right> \right) \frac{\partial \left( \Phi - \left< \Phi \right> \right)}{\partial \psi}
\right>,
 \end{equation}
using the radial derivative of the flux surface average of (\ref{eq:P}), and (\ref{eq:dnztdpsipot}).
Finally, inserting the flux function (\ref{eq:Kzbefsimpl}) back into (\ref{eq:Vzfinal}) and recalling that the radial variation of the magnetic field is negligible, the expression for the impurity flow when diamagnetic and radial flows are retained is obtained to be
 \begin{equation}
 \begin{aligned}  
\mathbf{V}_z
 = &
  \frac{\mathbf{B} }{n_z \left< B^2 \right>}
\left[
\frac{c I}{z_z e} \frac{\partial \left( \left< p_z \right> + P \right) }{\partial \psi}
 -  \frac{c I}{z_i e}  \left< n_z \right> T_i 
  \left(  
  \frac{d \ln  \left< p_i \right>}{d \psi} 
-\frac{3}{2} \frac{ d \ln T_i }{d \psi}
 \right)
 + u  \frac{\left< B^2 n_z \right>}{\left< n_i \right>}
   \right]
 \\ 
 &  
  - c R^2  \boldsymbol{\nabla} \zeta 
  \left( \frac{\partial \Phi}{\partial \psi} + \frac{1}{z_z e n_z} \frac{\partial p_z}{\partial \psi} \right)
  +
   \frac{cI }{B^2} 
\boldsymbol{\nabla} \zeta \times \boldsymbol{\nabla} \theta  
 \left( \frac{\partial \Phi}{\partial \theta} + \frac{1}{z_z e n_z} \frac{\partial p_z}{\partial \theta} \right) .
 \end{aligned}
 \label{eq:Vz}
\end{equation}
Notice that the last term has a radial component that is being retained.


\section{Parallel momentum \label{sec:parallelmom} } 

The parallel momentum equation (\ref{eq:||momKz}) can be further simplified by inserting $K_z$ from (\ref{eq:Kzbefsimpl}) 
and neglecting the radial variation of the magnetic field and $O \left( \epsilon \Delta \right)$ corrections
to obtain
\begin{equation}
\begin{aligned}
\mathbf{B} \cdot \boldsymbol{\nabla} P
 = 
 - M_i \left< \frac{\nu_{iz}}{n_z} \right> \left< n_i \right> 
 &
 \left\{
 \frac{c I T_i }{z_i e} 
  \left(  
  \frac{d \ln  \left< p_i \right>}{d \psi} 
-\frac{3}{2} \frac{ d \ln T_i }{d \psi}
 \right)
 \left( 
 n_z - \frac{B^2}{ \left< B^2 \right>}  \left< n_z \right>
 \right)
 \right.
 \\
& 
- u B^2  \left( \frac{n_z}{\left< n_i \right>} - \left< \frac{B^2}{\left< B^2 \right>} \frac{n_z}{\left< n_i \right>} \right>  \right)
 \\
& 
- \frac{c I}{z_z e}  \frac{\partial \left< p_z \right>}{\partial \psi} \left( 1 -  \frac{B^2}{ \left< B^2 \right>} \right) 
 \\
& 
\left.
+  \frac{c I}{z_z e} \frac{\partial \left[ P - \left( p_z - \left< p_z \right> \right) \right]}{\partial \psi} 
- c I n_z \frac{\partial \left(\Phi - \left< \Phi \right> \right)}{\partial \psi}
\right\}.
\end{aligned}
\label{eq:parmomdim}
\end{equation}
The left hand side can be evaluated to lowest order, by using (\ref{eq:P}) and recalling that the impurity density exhibits the strongest poloidal variation
 followed by the magnetic field, 
to find
\begin{eqnarray}
\frac{\partial P}{\partial \theta}
=
\left< T_z \right> \frac{\partial \left( n_z - \left< n_z \right> \right)}{\partial \theta}
\left(
1
 + \frac{\frac{z_z^2 \left< n_z \right>}{z_i^2 \left< n_i \right>}}{ 1 + \frac{\left< n_e \right>}{z_i \left< n_i \right>}  \frac{T_i}{z_i T_e}} \frac{T_i}{\left< T_z \right> }  \frac{n_z}{\left< n_z \right>}
\right).
\label{eq:dPdthetadnzdtheta}
\end{eqnarray}
Moreover, the dominant piece of the following term on the right hand side of (\ref{eq:parmomdim}) can be calculated by recalling that the impurity density exhibits both the strongest radial and poloidal variation and by using (\ref{eq:dnztdpsipot}) as follows:
\begin{equation}
\begin{aligned}
\frac{\partial \left[ P - \left( p_z - \left< p_z \right> \right) \right]}{\partial \psi} 
&
=
 \frac{z_z^2 T_i}{z_i^2 n_i \left( 1 + \frac{n_e}{z_i n_i} \frac{T_i}{z_i T_e} \right)} 
\left[ 
n_z \frac{\partial \left( n_z - \left< n_z \right> \right)}{\partial \psi} 
+ \left( n_z - \left< n_z \right> \right)  \frac{\partial \left< n_z \right>}{\partial \psi}  
\right] 
\\
&
=
z_z e n_z  \frac{\partial \left( \Phi - \left< \Phi \right> \right)}{\partial \psi}
+
 \frac{\frac{z_z^2 \left< n_z \right>}{z_i^2 \left< n_i \right>}}{ 1 + \frac{\left< n_e \right>}{z_i \left< n_i \right>} \frac{T_i}{z_i T_e} }
 \frac{T_i}{\left< T_z \right>} \left( \frac{n_z}{\left< n_z \right>} - 1 \right) \frac{\partial \left< p_z \right>}{\partial \psi} .
 \end{aligned}
 \label{eq:dPmpdpsi}
 \end{equation}
 
The conservation of parallel momentum equation for impurities in dimensionless form is found by combining (\ref{eq:parmomdim})-(\ref{eq:dPmpdpsi}) to obtain
 \begin{eqnarray}
 \left( 1 +  \alpha n \right)  \frac{\partial n}{\partial \vartheta}
 = 
 g \left( n - b^2 \right)
+ U b^2  \left( n - \left< n b^2 \right>  \right) 
+ D \left[ \alpha \left( n - 1 \right) +  b^2 - 1 \right];
\label{eq:parmomdPer}
\end{eqnarray}
where the dimensionless density, $n = \frac{n_z}{\left< n_z \right>} $, and magnetic field squared, $b^2 = \frac{B^2}{\left< B^2 \right>} $, present strong poloidal variation
which can be amplified by the following dimensionless flux functions 
\begin{equation}
 \alpha =  \frac{\frac{z_z^2  \left< n_z \right>}{z_i^2 \left< n_i \right>}}{1+ \frac{\left< n_e \right>}{z_i \left< n_i \right>} \frac{T_i}{z_i T_e}} \frac{T_i}{\left<T_z\right>}
 \sim 1,  
  \label{eq:alpha}
\end{equation}
\begin{equation}
 g = - \frac{c I}{z_i e} \frac{M_i \left< n_i \right>}{\left< \mathbf{B} \cdot \boldsymbol{\nabla} \theta \right>}
 \left< \frac{\nu_{iz}}{n_z} \right> 
 \frac{T_i}{\left< T_z \right>}
  \left(  
  \frac{d \ln  \left< p_i \right>}{d \psi} 
-\frac{3}{2} \frac{ d \ln T_i }{d \psi}
 \right),
\end{equation}
\begin{equation}
  U =  \frac{u \left< B^2 \right>}{\left< T_z \right>}   \frac{ M_i}{\left< \mathbf{B} \cdot \boldsymbol{\nabla} \theta \right> } \left< \frac{\nu_{iz}}{n_z} \right>
\end{equation}
and
\begin{equation}
  D = - \frac{c I}{z_z e} \frac{M_i \left< n_i \right>}{\left< \mathbf{B} \cdot \boldsymbol{\nabla} \theta \right>} \left< \frac{\nu_{iz}}{n_z} \right>  \frac{\partial \ln  \left< p_z \right> }{\partial \psi}.
  \label{eq:nondimvariables}
\end{equation}
Note that when the impurity diamagnetic effects are neglected, by taking the $D=0$ limit, Helander's equation (9) in~\cite{Helander1998} is recovered. 

The parallel momentum equation (\ref{eq:parmomdPer}) can be further simplified by neglecting all $O \left( \epsilon \Delta \right)$ corrections, to be consistent with previous assumption (\ref{eq:Binsidepolder}), to obtain
\begin{equation}
\begin{aligned}
 \left( 1 +  \alpha n \right)  \frac{\partial n}{\partial \vartheta}
& = 
 g \left( n - b^2 \right)
+ U  \left( n - 1  \right) 
+ D \left[ \alpha \left( n - 1 \right) +  b^2 - 1 \right]
\\
& = 
 \left( n - 1 \right) \left( g + U + \alpha D \right) 
 + \left( 1 - b^2 \right) \left( g - D \right)
 \end{aligned}
 \label{eq:parmomfinal}
\end{equation}


\section{Discussion and conclusions \label{sec:analysisp4}} 

Improving the modeling of impurities is expected to yield deeper insight into how to avoid impurity accumulation and a better understanding and diagnostic methods for H (and I) mode operation in a tokamak. 
In this section the expressions for the various components of the impurity flow are extended to include the two-dimensional impurity diamagnetic and radial flow effects.
These physical phenomena are shown here to obtain larger values of poloidal variation than the one-dimensional model~\cite{Helander1998}, as observed.
Finally, the novel expression for the impurity radial flux is derived and the diamagnetic effects are proven to beneficially enhance impurity removal, hence reducing or even preventing impurity accumulation while providing free fueling. 


\subsection{Poloidal impurity flow} 

The poloidal impurity flow has been experimentally observed~(see Figs. 4.4, 4.5, 4.6 and 4.7 of~\cite{Churchill2014phd}) to be much larger on the low field side of H-mode tokamak pedestals.
The novel diamagnetic and radial effects included in the prior sections result in an additional term in the poloidal impurity flow (\ref{eq:Vz}),
\begin{equation}  
\mathbf{V}_z \cdot \boldsymbol{\nabla} \theta
 =
  \frac{\mathbf{B} \cdot \boldsymbol{\nabla} \theta}{n_z \left< B^2 \right>}
\left[
 -  \frac{c I T_i}{z_i e}  \left< n_z \right>  
  \left(  
  \frac{d \ln  \left< p_i \right>}{d \psi} 
-\frac{3}{2} \frac{ d \ln T_i }{d \psi}
 \right)
 + u  \frac{\left< B^2 n_z \right>}{\left< n_i \right >}
 + \frac{c I}{z_z e} \frac{\partial \left( \left< p_z \right> + P \right) }{\partial \psi}
   \right],
\label{eq:Vzgradthetadisc}
\end{equation}
with respect to the one-dimensional model~\cite{Helander1998},
\begin{equation}  
\mathbf{V}_z \cdot \boldsymbol{\nabla} \theta
 =
  \frac{\mathbf{B} \cdot \boldsymbol{\nabla} \theta}{n_z \left< B^2 \right>}
\left[
 -  \frac{c I T_i}{z_i e}  \left< n_z \right>  
  \left(  
  \frac{d \ln  \left< p_i \right>}{d \psi} 
-\frac{3}{2} \frac{ d \ln T_i }{d \psi}
 \right)
 + u  \frac{\left< B^2 n_z \right>}{\left< n_i \right >}
 \right].
 \label{eq:VzgradthetadiscPer}
 \end{equation} 
Both sources of poloidal variation are located out front of the square bracket in (\ref{eq:VzgradthetadiscPer}): the poloidal magnetic field, whose poloidal variation is weak in the large aspect ratio limit, and the inverse of the impurity density, which drives the poloidal impurity flow to be significantly larger on the outboard as measured.
The last term in~(\ref{eq:Vzgradthetadisc}), $\frac{\partial \left( \left< p_z \right> + P \right) }{\partial \psi}$, introduces poloidal variation within the square bracket.

The poloidal variation of the last term in (\ref{eq:Vzgradthetadisc}) can be made more explicit by using (\ref{eq:dPmpdpsi}) and recalling that the impurity density contains the strongest radial and poloidal variation, so that
\begin{equation}  
\frac{\partial \left( \left< p_z \right> + P \right) }{\partial \psi}
=
\left( 
\left< T_z \right> + \alpha T_i \frac{n_z}{\left< n_z \right>} 
\right) \frac{\partial n_z}{\partial \psi}
- \alpha T_i  \frac{\partial \left< n_z \right>}{\partial \psi} .
\label{eq:dpdpsiVzpol} 
\end{equation}
By substituting (\ref{eq:dpdpsiVzpol}) into (\ref{eq:Vzgradthetadisc}) and identifying the non-dimensional flux functions defined in (\ref{eq:alpha})-(\ref{eq:nondimvariables}), it is shown that the poloidal impurity flux over the poloidal magnetic field,
\begin{equation}
\frac{ n_z \mathbf{V}_z \cdot \boldsymbol{\nabla} \theta }{ \mathbf{B} \cdot \boldsymbol{\nabla} \theta }
 =  
 \frac{ \left< n_z \right> \left< \mathbf{B} \cdot \boldsymbol{\nabla} \theta \right>}{\left< B^2 \right> M_i \left< n_i \right> \left< \frac{\nu_{iz}}{n_z} \right>}
\left( g+U+\alpha D \right)
+ 
\frac{c I}{z_z e \left< B^2 \right>}
\left( 
1 + \alpha \frac{T_i}{\left< T_z \right>} \frac{n_z}{\left< n_z \right>} 
\right) \frac{\partial n_z}{\partial \psi} ,
\label{eq:Vzpolnewpolvar}
\end{equation}
 is not a flux function in contrast to~\cite{Helander1998}.
The additional poloidal variation introduced by the diamagnetic and radial flow effects is given by the last term on the right hand side of (\ref{eq:Vzpolnewpolvar}).
As the right hand side of Fig.~1 in \cite{Churchill2015} shows, the impurity density is larger on the high field side and its radial gradient is steeper (more negative) on the high field side. 
Therefore the final ($\frac{\partial n_z}{\partial \psi}$) term makes the poloidal flow asymmetry larger than previous models~\cite{Helander1998}.

It is worth noticing that the combination $g+U+\alpha D$ in (\ref{eq:Vzpolnewpolvar}) must be positive when the poloidal flow is positive, since the second term in the right hand side of (\ref{eq:Vzpolnewpolvar}) is negative. 
In addition, by rewriting (\ref{eq:dpdpsiVzpol}) as 
\begin{eqnarray}  
\frac{\partial \left( \left< p_z \right> + P \right) }{\partial \psi}
=
\left[ 1 + \alpha \left( n - 1 \right) \right] \left< T_z \right> \frac{\partial \left< n_z \right>}{\partial \psi} 
+ \left( 1 + \alpha n \right) \left< T_z \right> \frac{\partial \left( n_z - \left< n_z \right> \right) }{\partial \psi} ,
\label{eq:dpdpsiVzpolnd}
\end{eqnarray}
it is seen that this quantity is negative on the high field side for inboard accumulation and more negative inboard impurity density slope in agreement with H-mode experiments~\cite{Theiler2014,Churchill2015}.
It thus follows from (\ref{eq:Vzgradthetadisc}) that $g+U$ should be positive for the poloidal flow to be positive on the high field side.

\subsection{Parallel impurity flow} 

The parallel impurity flow has also been measured~(see Figs. 4.4, 4.5, 4.6 and 4.7 of~\cite{Churchill2014phd}) to be much larger on the outboard side of H-mode tokamak pedestals.
The novel diamagnetic and radial effects and the stronger poloidal variation of the radial electric field included in the model also lead to an  additional term in the parallel impurity flow (\ref{eq:Vz}),
 \begin{multline}
V_{z\parallel}
 = 
 - \frac{c I}{B} \frac{\partial \left< \Phi \right>}{\partial \psi}
 +
  \frac{B}{n_z \left< B^2 \right>}
\left[
 -  \frac{c I}{z_i e}  \left< n_z \right> T_i 
  \left(  
  \frac{d \ln  \left< p_i \right>}{d \psi} 
-\frac{3}{2} \frac{ d \ln T_i }{d \psi}
 \right)
 + u  \frac{\left< B^2 n_z \right>}{\left< n_i \right>}
   \right]
 \\ 
+ \frac{c I}{z_z e n_z B}
\left\{
\frac{B^2}{ \left< B^2 \right>} \frac{\partial \left( \left< p_z \right> + P \right) }{\partial \psi} 
  - \left[ \frac{\partial p_z}{\partial \psi} + z_z e n_z \frac{\partial \left( \Phi - \left< \Phi \right> \right)}{\partial \psi} \right]
  \right\},
  \label{eq:Vzparanalysis}
\end{multline}
with respect to the one-dimensional model~\cite{Helander1998} which contains only the terms in the first line of (\ref{eq:Vzparanalysis}).
This term introduces a new source poloidal variation in addition to previous drives for the parallel impurity flow to be significantly larger on the low field side as experimentally observed, given by the inverse of the magnetic field on the first term in the first line of (\ref{eq:Vzparanalysis}) and the poloidal variation of the inverse of the impurity density dominating that given by the magnetic field on the second term.

The term in (\ref{eq:Vzparanalysis}) containing diamagnetic and radial flow effects and the poloidal variation of the radial electric field is evaluated by using (\ref{eq:dPmpdpsi}) to find
\begin{multline}  
\frac{B^2}{ \left< B^2 \right>} \frac{\partial \left( \left< p_z \right> + P \right) }{\partial \psi} 
- \left[ \frac{\partial p_z}{\partial \psi}  + z_z e n_z \frac{\partial \left( \Phi - \left< \Phi \right> \right)}{\partial \psi} \right]
 = 
 \\
=
 \frac{\partial \left< p_z \right> }{\partial \psi}  
 \left[
\left( \frac{B^2}{ \left< B^2 \right>}  - 1  \right)
+
\alpha
 \frac{T_i}{\left< T_z \right>} \left( \frac{n_z}{\left< n_z \right>} - 1 \right) 
 \frac{B^2}{ \left< B^2 \right>}
 \right];
 \label{eq:Vzparnewterm}
\end{multline}
where to be consistent with (\ref{eq:Binsidepolder}) the following terms have been neglected:
\begin{equation}
\frac{ \left( \frac{B^2}{ \left< B^2 \right>}  - 1 \right) \frac{\partial \left( p_z - \left< p_z \right> \right) }{\partial \psi} }
{ z_z e n_z \frac{\partial \left< \Phi \right>}{\partial \psi}}
\sim
\frac{ 
z_z e n_z  \left( \frac{B^2}{ \left< B^2 \right>}  - 1 \right)
 \frac{\partial \left( \Phi - \left< \Phi \right> \right)}{\partial \psi}}
{ z_z e n_z \frac{\partial \left< \Phi \right>}{\partial \psi}}
\sim
\epsilon \Delta
\ll 1.
\end{equation}
From (\ref{eq:Vzparnewterm}), it can be seen that the new term is negative on the high field side and positive on the low field side, since the impurity density is measured~\cite{Churchill2015} to be larger on the high field side in H-mode.
Consequently, the two-dimensional model with diamagnetic and radial effects that 
allows stronger poloidal variation of the radial electric field  can capture substantially stronger in-out asymmetries in the positive parallel impurity flow than the previous state-of-the-art models~\cite{Helander1998}. 

\subsection{Radial impurity particle flux} 

By taking the toroidal projection of impurity momentum conservation (\ref{eq:momconsvimp}) and using axisymmetry, the radial particle flux is found to be
\begin{eqnarray}
n_z \mathbf{V}_z \cdot \boldsymbol{\nabla} \psi
= 
\frac{c}{z_z e}
R^2 \boldsymbol{\nabla}  \zeta 
\cdot
\left(
\mathbf{R}_{iz}
+ M_z n_z \mathbf{V}_z \cdot \boldsymbol{\nabla} \mathbf{V}_z
+ \boldsymbol{\nabla} \cdot \boldsymbol{\pi}_z
\right).
\label{eq:tormomcons}
\end{eqnarray}
Using the estimates (\ref{eq:neglinertia||mom}), (\ref{eq:neglviscd||mom}), (\ref{eq:neglviscod||mom}), (\ref{eq:inertiaradnegl}), (\ref{eq:perpmomfricrad}) and (\ref{eq:viscradnegl})
to neglect small terms leads to the conclusion
that friction dominates on the right hand side of (\ref{eq:tormomcons}), 
thus satisfying ambipolarity since
\begin{eqnarray}
z_z n_z \mathbf{V}_z \cdot \boldsymbol{\nabla} \psi
=
\frac{c}{e} R^2 \boldsymbol{\nabla}  \zeta 
\cdot \mathbf{R}_{iz}
=
- \frac{c}{e} R^2 \boldsymbol{\nabla}  \zeta 
\cdot \mathbf{R}_{zi}
=
- z_i n_i \mathbf{V}_i \cdot \boldsymbol{\nabla} \psi.
\label{eq:radfluxfricamb}
\end{eqnarray}

The calculation of the radial impurity particle flux can be further simplified by noticing that the parallel friction dominates since
\begin{eqnarray}
\frac{\mathbf{R}_{iz\perp} \cdot R \boldsymbol{\nabla}  \zeta}
{\mathbf{R}_{iz\parallel} \cdot R \boldsymbol{\nabla}  \zeta }
=
\frac{- \mathbf{R}_{iz} \cdot  \frac{\mathbf{B} \times \boldsymbol{\nabla} \psi }{B R B_p} \frac{B_p}{B}}
{R_{iz\parallel} \frac{I}{R B} }
\sim
\frac{B_p^2}{B^2}
\ll 1,
\end{eqnarray}
where the estimate (\ref{eq:Rizpolord}) has been used.
As a result, the flux-surface averaged radial impurity particle flux is given to lowest order by
\begin{equation}
\begin{aligned}
\left< n_z \mathbf{V}_z \cdot \boldsymbol{\nabla} \psi \right>
& =
- \frac{cI}{z_z e}
\left< \frac{R_{zi\parallel}}{B} \right>
 =
- \frac{cI}{z_z e \left< B^2 \right>}
\left< \frac{B R_{zi\parallel}}{1 + \left( b^2 - 1 \right)} \right>
 \\
& =
\frac{cI}{z_z e \left< B^2 \right>}
\left< \frac{B R_{zi\parallel} \left[ \left( b^2 - 1 \right) - 1 \right] }{1 - \left( b^2 - 1 \right)^2} \right>
\\
& =
\frac{cI}{z_z e \left< B^2 \right>}
\left< B R_{zi\parallel} \left( b^2 - 1 \right) \right>
+ O \left( \Delta \epsilon^2  \frac{\rho_{pz}}{qR} n_z v_{Tz} R B_p \right);
\end{aligned}
\end{equation}
where in the last form the denominator has been Taylor expanded and the solubility constraint, $\left< B R_{zi\parallel} \right> = 0$, has been used. 
Finally, substituting the lowest order expression for the friction, whole poloidal variation is proportional to the right hand side of (\ref{eq:parmomfinal}) divided by the magnetic field magnitude, the radial impurity flux becomes
\begin{eqnarray}
\frac{ \left< n_z \mathbf{V}_z \cdot \boldsymbol{\nabla} \psi \right> }
{ \frac{cI \left< \mathbf{B} \cdot \boldsymbol{\nabla} \theta \right> \left< p_z \right>}{z_z e \left< B^2 \right>} }
=
 \left< \left( n - 1 \right) \left( b^2 - 1\right) \right> \left( g + U + \alpha D \right) 
 -  \left< \left( 1 - b^2 \right)^2 \right> \left( g - D \right).
 \label{eq:partfluxnd}
\end{eqnarray}
Note that (13) in \cite{Helander1998} is correctly recovered for $D=0$ to lower order.

For illustrative purposes, a first-order cosinusoidal poloidal variation is considered for the dimensionless magnetic field, $b^2 = 1 - 2 \epsilon \cos \vartheta$, along with a first-order Fourier profile for the dimensionless impurity density, with both a cosinusoidal and sinusoidal term in order to allow for in-out and up-down asymmetries. 
It can then be observed that both of the flux surface averages on the right hand side of (\ref{eq:partfluxnd}) are positive, since both the magnetic field and the impurity density are larger on the inboard than on the outboard.
Since $D$ is positive by definition and it is always proportional to a positive coefficient in (\ref{eq:partfluxnd}), more impurities go out and ions go in as $D$ increases. 
In conclusion, being in a regime with large diamagnetic drift helps to remove impurities. 

\appendix

\section{Collision frequencies\label{collisionfreq}}

The collisional frequencies between impurities and/or main ions are given~\cite{Helander2005} by
\begin{equation*}
\begin{aligned}
\nu_{ii} =& \frac{4 \sqrt{\pi} z_i^4 e^4 n_i \ln \Lambda}{3 M_i^{\frac{1}{2}} T_i^{\frac{3}{2}}} \\
\nu_{iz} = &\frac{4 \sqrt{2 \pi} z_i^2 z_z^2 e^4 n_z \ln \Lambda}{3 M_i^{\frac{1}{2}} T_i^{\frac{3}{2}}} \\
\nu_{zz} =& \frac{4 \sqrt{\pi} z_z^4 e^4 n_z \ln \Lambda}{3 M_z^{\frac{1}{2}} T_z^{\frac{3}{2}}}
\end{aligned}
\end{equation*}
and
\begin{equation*}
\begin{aligned}
\nu_{zi} = & \frac{8 \sqrt{2 \pi} M_i^{\frac{1}{2}} z_i^2 z_z^2 e^4 n_i \ln \Lambda}{3 M_z T_i^{\frac{3}{2}}},
\end{aligned}
\end{equation*}
where $\nu_{12}$ denotes the collisional frequency of species $1$ with $2$ and $\ln \Lambda$ is the Coulomb logarithm.
Note that the collision frequencies between impurities and main ions satisfy $2 M_i n_i {\nu}_{iz} = M_z n_z {\nu}_{zi}$.

The sizes of the collisional frequencies can be compared to find\begin{equation*}
\begin{aligned}
\frac{\nu_{iz}}{\nu_{ii}}
 = & \sqrt{2} z_{\mbox{eff}}
&& \sim  z_{\mbox{eff}},
\\
\frac{\nu_{zz}}{\nu_{zi}} = &
\frac{1}{2\sqrt{2}} \left( \frac{T_i}{T_z} \right)^{\frac{3}{2}} \left( \frac{M_z}{M_i} \right)^{\frac{1}{2}} z_{\mbox{eff}} 
&& \sim \left( \frac{z_z}{z_i} \right)^{\frac{1}{2}} z_{\mbox{eff}} 
\end{aligned}
\end{equation*}
and
\begin{equation*}
\begin{aligned}
\frac{\nu_{zz}}{\nu_{ii}} = &
 \sqrt{2} \left( \frac{z_z}{z_i} \right)^2 \left( \frac{M_i}{M_z} \right)^{\frac{1}{2}} z_{\mbox{eff}} 
 && \sim \left( \frac{z_z}{z_i} \right)^{\frac{3}{2}}  z_{\mbox{eff}} ;
\end{aligned}
\end{equation*}
where $z_{\mbox{eff}} = \frac{z_z^2 n_z}{z_i^2 n_i}$ is the effective charge of the impurities.

\section{Maxwellian impurity distribution function to lowest order \label{sec:paper4_a3}}

In order for the impurity distribution function to be a drifting Maxwellian to lowest order,
\begin{eqnarray}
f_{zM}
= 
n_z \left( \frac{M_z}{2 \pi T_z} \right)^{\frac{3}{2}}
\exp \left( - \frac{M_z w_z^2 }{2 T_z} \right),
\end{eqnarray}
its first-order correction $f_{z1}$ should be much smaller.
In this appendix, the implied parameter requirements 
are deduced from the static Fokker-Planck equation in spatial and relative velocity, $\mathbf{w}_z = \mathbf{v}_z - \mathbf{V}_z$, variables \cite{Catto2004}:
\begin{equation}
\mathbf{v}_z \cdot \boldsymbol{\nabla} f_z
+ \left[ 
\Omega_z \mathbf{w}_z \times \frac{\mathbf{B}}{B} 
+ \frac{z_z e}{M_z} 
\left( 
\frac{\mathbf{V}_z \times \mathbf{B}}{c}  - \boldsymbol{\nabla} \Phi
\right)
- \mathbf{v}_z \cdot \boldsymbol{\nabla} \mathbf{V}_z
\right] 
\cdot 
\boldsymbol{\nabla}_{w_z}  f_z
= C_{zz} 
+ C_{zi};
\label{eq:kinw}
\end{equation}
where the gradient in relative velocity space is given by 
\begin{eqnarray}
\mathbf{\nabla}_{w_z} 
= \frac{\mathbf{B}}{B} \frac{\partial}{\partial w_{z \parallel}}
+ \frac{\mathbf{w}_{z \perp}}{w_{z \perp}} \frac{\partial}{\partial w_{z \perp}}
- \frac{1}{w_{z \perp}^2} \mathbf{w}_z \times \frac{\mathbf{B}}{B} \frac{\partial}{\partial {\varphi}_z}
\label{eq:derw}
\end{eqnarray} 
with the gyrophase ${\varphi}_z$ defined by $\frac{\mathbf{w}_{z \perp}}{w_{z \perp}} = \frac{\boldsymbol{\nabla} \psi}{R B_p} \cos {\varphi}_z + \frac{\mathbf{B} \times\boldsymbol{\nabla} \psi}{B R B_p} \sin {\varphi}_z $.
It is convenient to decompose the first-order distribution function into its gyroaverage, $\bar{f}_{z1} = \left< f_{z1} \right>_{{\varphi}_z} = \frac{1}{2 \pi}  \oint d {\varphi}_z f_{z1}$, and gyrophase dependent part, $\tilde{f}_{z1} = f_{z1} - \bar{f}_{z1}$.

\subsection{Gyrophase independent first-order correction:} 

The gyroaveraged first-order kinetic equation for the impurities is thus
\begin{eqnarray}
\mathbf{w}_{z\parallel} \cdot \boldsymbol{\nabla} f_{zM}
- \frac{z_z e}{M_z} \frac{\mathbf{B}}{B} \cdot \boldsymbol{\nabla} \Phi  \frac{\partial f_{zM}}{\partial w_{z\parallel}}
=
C_{zz1} \left\{ \bar{f}_{z1} \right\}
+
\left< C_{zi1} \left\{ f_{i1} \right\} \right>_{{\varphi}_z};
\label{eq:impkinga}
\end{eqnarray}
since the other left hand side terms are negligible:
\begin{equation}
\frac{
\mathbf{V}_z \cdot \boldsymbol{\nabla} f_{zM}
}{
\mathbf{w}_{z\parallel} \cdot \boldsymbol{\nabla} f_{zM}
}
 \sim
\frac{
-\frac{w_{z\perp}}{2} \frac{\partial f_{zM}}{\partial w_{z\perp}}
\boldsymbol{\nabla} \cdot \mathbf{V}_z
}{
\mathbf{w}_{z\parallel} \cdot \boldsymbol{\nabla} f_{zM}
}
\sim
\frac{
\left(
\frac{w_{z\perp}}{2} \frac{\partial f_{zM}}{\partial w_{z\perp}}
- w_{z\parallel} \frac{\partial f_{zM}}{\partial w_{z\parallel}}
\right)
\frac{\mathbf{B}}{B} \cdot \boldsymbol{\nabla} \mathbf{V}_z \cdot \frac{\mathbf{B}}{B} 
}{
\mathbf{w}_{z\parallel} \cdot \boldsymbol{\nabla} f_{zM}
}
\sim
\frac{\rho_{pz}}{L_{n_z}}
\ll 1,
\end{equation}
and
\begin{eqnarray}
 \frac{
- \mathbf{V}_z \cdot \boldsymbol{\nabla} \mathbf{V}_z \cdot \frac{\mathbf{B}}{B} \frac{\partial f_{zM}}{\partial w_{z\parallel}}
}{
\mathbf{w}_{z\parallel} \cdot \boldsymbol{\nabla} f_{zM}
}
\sim
\frac{\rho_{pz}^2}{L_{n_z}^2}
\ll 1.
\end{eqnarray}
So as to perform the gyroaverage, it has been used that $\left< \mathbf{w}_{z\perp} \mathbf{w}_{z\perp} \right> = \frac{w_{z\perp}^2}{2} \left( \mathbf{I} - \frac{\mathbf{B}}{B} \frac{\mathbf{B}}{B} \right)$ and that $\mathbf{I} : \boldsymbol{\nabla} \mathbf{V}_z = \boldsymbol{\nabla} \cdot \mathbf{V}_z$, where $\mathbf{I}$ is the identity matrix.
In order to calculate the estimates, it is worth recalling that the impurity density presents the strongest poloidal and poloidal variation.
In addition, the relative velocity is of the order of the impurity thermal velocity in all directions.

If the parallel streaming and radial electric field terms balance self-collisions in (\ref{eq:impkinga}), the size of the gyrophase independent first-order correction to the lowest-order impurity distribution function is given by
\begin{eqnarray}
\frac{\bar{f}_{z1}}{f_{zM}} 
\sim
\Delta \frac{v_{Tz}}{\nu_{zz} qR} 
\sim
\Delta \frac{\lambda_z}{qR} 
\ll 1;
\label{eq:f1gacollfreq}
\end{eqnarray}
while if unlike collisions balance self-collisions then
\begin{eqnarray}
\frac{\bar{f}_{z1}}{f_{zM}} 
\sim
\frac{\nu_{zi}}{\nu_{zz}}
\sim
\sqrt{\frac{z_i}{z_z}}
\ll 1,
\end{eqnarray}
which justifies the highly charged impurity assumption.

\subsection{Gyrophase dependent first order correction:} 

Subtracting from the Fokker-Planck equation (\ref{eq:kinw}) its flux surface average,
the gyrophase dependent first-order kinetic equation for the impurities is thus given by
\begin{eqnarray}
\mathbf{w}_{z\perp} \cdot \boldsymbol{\nabla} f_{zM}
+ \frac{\boldsymbol{\nabla} p_z }{M_z n_z} \cdot  \frac{\mathbf{w}_{z\perp}}{w_{z\perp}} \frac{\partial f_{zM}}{\partial w_{z\perp}}
- \Omega_z \frac{\partial \tilde{f}_{z1}}{\partial {\varphi}_z } 
=
C_{zi1} \left\{ f_{i1} \right\} 
- \left< C_{zi1} \left\{ f_{i1} \right\}  \right>_{{\varphi}_z};
\label{eq:impkingd}
\end{eqnarray}
where the following terms have been neglected on the left hand side:
\begin{eqnarray}
\frac{ - \mathbf{w}_{z\perp} \cdot \boldsymbol{\nabla} \mathbf{V}_z 
\cdot \frac{\mathbf{B}}{B} \frac{\partial f_{zM}}{\partial w_{z\parallel}}
}{\mathbf{w}_{z\perp} \cdot \boldsymbol{\nabla} f_{zM}}
\sim
\frac{\rho_{pz}}{L_{n_z}}
\ll 1
\end{eqnarray}
and
\begin{eqnarray}
\frac{
\frac{1}{w_{z\perp}} \frac{\partial f_{zM}}{\partial w_{z\perp}}
\left[
\frac{w_{z\perp}^2}{2} 
\left(
\mathbf{I}
- \frac{\mathbf{B}}{B} \frac{\mathbf{B}}{B}
\right)
- 
\mathbf{w}_{z\perp} \mathbf{w}_z
\right]
: \boldsymbol{\nabla} \mathbf{V}_z
}{\mathbf{w}_{z\perp} \cdot \boldsymbol{\nabla} f_{zM}}
\sim
\frac{\rho_z}{L_{n_z}}
\ll 1,
 \end{eqnarray}
with $\mathbf{a}_1 \mathbf{a}_2 : \boldsymbol{\nabla} \mathbf{V}_z = \mathbf{a}_2 \cdot \boldsymbol{\nabla} \mathbf{V}_z \cdot \mathbf{a}_1$ for generic vectors $\mathbf{a}_1$ and $\mathbf{a}_2$.
Moreover, the perpendicular momentum conservation is assumed in (\ref{eq:Vzperp}) to be dominated by the Lorentz, electrostatic and isotropic pressure forces:
 \begin{eqnarray}
\frac{
\frac{\boldsymbol{\nabla} \cdot \boldsymbol{\pi}_z - \mathbf{R}_{zi}}{M_z n_z}
\cdot  \frac{\mathbf{w}_{z\perp}}{w_{z\perp}} \frac{\partial f_{zM}}{\partial w_{z\perp}}
}{\mathbf{w}_{z\perp} \cdot \boldsymbol{\nabla} f_{zM}}
\sim
\frac{
\frac{\boldsymbol{\nabla} \cdot \boldsymbol{\pi}_z - \mathbf{R}_{zi}}{M_z n_z}
\cdot  \frac{\mathbf{w}_{z\perp}}{w_{z\perp}} \frac{\partial f_{zM}}{\partial w_{z\perp}}
}{\frac{\boldsymbol{\nabla} p_z}{M_z n_z}
\cdot  \frac{\mathbf{w}_{z\perp}}{w_{z\perp}} \frac{\partial f_{zM}}{\partial w_{z\perp}}}
\ll 1.
\end{eqnarray}
Finally, the term containing the gyrofrequency overtakes the like collision operator since
\begin{eqnarray}
 \frac{
C_{zz1} \left\{ \tilde{f}_{z1} \right\}
}{
\Omega_z  \frac{\partial \tilde{f}_{z1}}{\partial \varphi_z}
}
\sim
\frac{
\nu_{zz}
}{
\Omega_z
}
\ll 1.
\end{eqnarray}

If the term involving the gyrofrequency competes with the perpendicular streaming in (\ref{eq:impkingd}),
the size of the gyrophase dependent first-order correction to the impurity distribution function is given by
\begin{eqnarray}
\frac{\tilde{f}_{z1}}{f_{zM}}
\sim 
\frac{v_{Tz}}{\Omega_z L_{n_z}}
\sim
\frac{\rho_z}{L_{n_z}}
\ll 1,
\end{eqnarray}
while if the former balances unlike collisions then
\begin{eqnarray}
\frac{\tilde{f}_{z1}}{f_{zM}}
\sim 
\frac{\nu_{zi}}{\Omega_z} 
\frac{f_{iM}}{f_{zM}}
\frac{f_{i1}-\left< f_{i1} \right>}{f_{iM}}
\sim 
\frac{\nu_{zz}}{\Omega_z} 
\frac{f_{i1}-\left< f_{i1} \right>}{f_{iM}}
\sim 
\frac{\rho_z}{\lambda_z} 
\frac{f_{i1}-\left< f_{i1} \right>}{f_{iM}}
\ll 1;
\end{eqnarray}
where it has been used that $\frac{f_{iM}}{f_{zM}}\sim \sqrt{\frac{z_z}{z_i}}\sim \frac{\nu_{zz}}{\nu_{zi}}$ for the orderings in hand.
This is satisfied automatically since $\rho_z \ll \lambda_z$.

\section{Neglecting heat flux and viscous contributions to the parallel momentum and energy conservation equations}

In this appendix, it is justified that the divergence of the heat flux does not significantly affect the impurity energy balance (\ref{eq:neglheatenergy}) for the new orderings.
Furthermore, the anisotropic force and the viscous dissipation are proven to be negligible contributions to the impurity parallel momentum (\ref{eq:neglviscd||mom}, \ref{eq:neglviscod||mom}) and energy (\ref{eq:neglviscdenergy}, \ref{eq:neglviscodenergy}) conservation respectively,
by using the calculated impurity flux (\ref{eq:Vz}).

Although there may be additional components coming from the presence of unlike collisions on the impurity Fokker-Planck equation, the most complete expressions for the collisional heat flux and viscous tensor to date~\cite{Catto2004} are used for the estimates.

\subsection{Heat flux} 

The collisional and diamagnetic heat flux obtained in Eq. 39 of \cite{Catto2004} ($C$), 
\begin{eqnarray}
\mathbf{q}_{zC} 
= 
\frac{p_z}{M_z}
\left(
- \frac{125}{32 \nu_{zz}} \frac{\mathbf{B}}{B} \frac{\mathbf{B}}{B} \cdot \boldsymbol{\nabla} T_z
+ \frac{5 }{2  \Omega_z} \frac{\mathbf{B}}{B} \times \boldsymbol{\nabla} T_z
- \frac{2 \nu_{zz}}{\Omega^2_z} \boldsymbol{\nabla}_{\perp} T_z
\right),
\end{eqnarray}
has the following size in the parallel, poloidal and radial directions:
\begin{equation}
\begin{aligned}
q_{z\parallel C} 
 & \sim 
- \frac{p_z T_z}{M_z \nu_{zz}}
\frac{\mathbf{B} \cdot \boldsymbol{\nabla} \theta}{B} \frac{\partial \ln T_z}{\partial \theta}
& & \sim
p_z v_{Tz} \frac{\lambda_z}{qR} \frac{\partial \ln T_z}{\partial \theta},
\\
 \mathbf{q}_{zC} \cdot \frac{\mathbf{B} \times \boldsymbol{\nabla} \psi}{B R B_p}
 &  \sim
\frac{p_z R B_p}{M_z \Omega_z}
 \left( 
 \frac{\partial T_z}{\partial \psi}
+
\frac{\boldsymbol{\nabla} \theta \cdot \boldsymbol{\nabla} \psi}{R^2 B_p^2} \frac{\partial T_z}{\partial \theta}
\right)
& &  \sim
p_z v_{Tz} \frac{z_i}{z_z} \frac{\rho_z}{L_{n_z}},
\\
  \mathbf{q}_{zC} \cdot \frac{\boldsymbol{\nabla} \psi}{R B_p}
  & \sim
\frac{p_z T_z}{M_z \Omega_z } \frac{\mathbf{B} \times \boldsymbol{\nabla} \theta \cdot \boldsymbol{\nabla} \psi}{B R B_p}  \frac{\partial \ln T_z}{\partial \theta}
& & \sim
p_z v_{Tz} \frac{\rho_{pz}}{qR} \frac{\partial \ln T_z}{\partial \theta}.
\end{aligned}
\end{equation}
Due to axisymmetry, the size of the divergence of the heat flux is thus given by
\begin{multline}
\boldsymbol{\nabla} \cdot \mathbf{q}_{zC}
 = 
\mathbf{B} \cdot \boldsymbol{\nabla} \theta \left[
\frac{\partial}{\partial \theta} \left( \frac{\mathbf{q}_{zC} \cdot \boldsymbol{\nabla} \theta}{\mathbf{B} \cdot \boldsymbol{\nabla} \theta} \right)
+ \frac{\partial}{\partial \psi} \left( \frac{\mathbf{q}_{z\perp C} \cdot \boldsymbol{\nabla} \psi}{\mathbf{B} \cdot \boldsymbol{\nabla} \theta} \right) 
\right]
\\
 \sim
\frac{p_z v_{Tz}}{qR}
\max \left\{
 \Delta \frac{\lambda_z}{qR} \frac{\partial \ln T_z}{\partial \theta},
\Delta  \frac{z_i}{z_z} \frac{\rho_{pz}}{L_{n_z}},
\frac{\rho_{pz}}{L_{n_z}} \frac{\partial \ln T_z}{\partial \theta}
\right\},
\label{eq:divqzC}
\end{multline}
where (\ref{eq:Bxthetapsi}) and the fact that the strongest radial and poloidal variation are exhibited by the impurity density have been used. 

\subsection{Viscous tensor \label{sec:viscoustensorapp}} 


\paragraph{Diagonal:}
The diagonal ($d$) part of the viscous tensor is obtained on Eq. (42) of \cite{Catto2004}:
\begin{equation}
\begin{aligned}
\boldsymbol{\pi}_{zdC} \cdot \frac{\mathbf{B}}{B}
= 
\frac{2}{3} \frac{\mathbf{B}}{B} 
&
\left\{
\frac{M_z}{p_z T_z} \left( 0.412 q^2_{z\parallel C} - 0.064 q_{zC}^2 \right) 
\hspace{-10.0cm}
\right.&
 \\
& 
+ \frac{0.960}{\nu_{zz}} \left( \mathbf{I} - 3\frac{\mathbf{B}}{B} \frac{\mathbf{B}}{B} \right) 
\hspace{-0.1cm}
: \hspace{-0.4cm}
& \left[
0.246 \left( \boldsymbol{\nabla} \mathbf{q}_{zC} - \mathbf{q}_{zC} \boldsymbol{\nabla} \ln p_z + \frac{4}{15} \boldsymbol{\nabla} \mathbf{q}_{z\parallel C} \right)
\right.
 \\
& &
\left.
\left.
+ \left( p_z \boldsymbol{\nabla} \mathbf{V}_z + \frac{2}{5} \boldsymbol{\nabla} \mathbf{q}_{zC} \right)
\right] 
 \right\}.
 \end{aligned}
  \end{equation}
It contains heat flux (\ref{eq:divqzC}) and impurity flux (\ref{eq:Vzgradord}) terms, whose size ratios are given by 
 \begin{equation}
 \frac{ \frac{M_z}{p_z T_z} q_{zC}^2}
{ \frac{p_z}{\nu_{zz}}  \boldsymbol{\nabla} \cdot \mathbf{V}_z}
 \sim
\frac{1}{\Delta} 
\max \left\{
\frac{\lambda_z}{qR} \frac{L_{n_z}}{\rho_{pz}}  \left(  \frac{\partial \ln T_z}{\partial \theta}  \right)^2,
\frac{qR}{\lambda_z} \frac{\rho_{pz}}{L_{n_z}} \frac{z_i^2}{z_z^2} \frac{B_p^2}{B^2}  ,
 \frac{L_{n_z}}{\lambda_z} \frac{\rho_{pz}}{qR} \left( \frac{\partial \ln T_z}{\partial \theta}  \right)^2
\right\}
\end{equation}
and
\begin{equation}
 \frac{ \frac{1}{\nu_{zz}} \boldsymbol{\nabla} \cdot \mathbf{q}_{zC}}
{ \frac{p_z}{\nu_{zz}}  \boldsymbol{\nabla} \cdot \mathbf{V}_z}
 \sim  
\max \left\{
\frac{\lambda_z}{qR}  \frac{L_{n_z}}{ \rho_{pz}} \frac{\partial \ln T_z}{\partial \theta},
\frac{z_i}{z_z}, 
\frac{1}{\Delta}  \frac{\partial \ln T_z}{\partial \theta}
 \right\}.
\end{equation}
The size of the viscous tensor diagonal term, 
for Maxwell-Boltzmann main ions (\ref{eq:ionMBconst}), is thus given by
 \begin{equation}
 \frac{\frac{\mathbf{B}}{B}   \cdot \boldsymbol{\pi}_{zdC} \cdot \frac{\mathbf{B}}{B}}
{ \frac{p_z}{\nu_{zz}}  \boldsymbol{\nabla} \cdot \mathbf{V}_z}
\sim
\max \left\{
1,
\frac{\lambda_z}{qR}  \frac{L_{n_z}}{ \rho_{pz}} \frac{\partial \ln T_z}{\partial \theta} ,
\frac{1}{\Delta} \frac{L_{n_z}}{\lambda_z} \frac{\rho_{pz}}{qR} \left( \frac{\partial \ln T_z}{\partial \theta}  \right)^2
\right\},
\label{eq:pizdb}
\end{equation}
where the size of the divergence of the impurity flux is calculated on (\ref{eq:Vzgradord}).

The contribution of the diagonal viscous tensor to the viscous force is thus indeed negligible compared to the pressure and potential gradients in the \emph{parallel momentum} equation,
\begin{equation}
\frac{
\boldsymbol{\nabla} \cdot \boldsymbol{\pi}_{zdC} \cdot \frac{\mathbf{B}}{B}
}{ {\nabla}_{\parallel} p_z + z_z e n_z {\nabla}_{\parallel} \Phi }
\sim
\max \left\{
\frac{\rho_{pz}}{L_{n_z}} \Delta \frac{\lambda_z}{qR},
 \left( \Delta \frac{\lambda_z}{qR} \right)^2 \frac{1}{\Delta} \frac{\partial \ln T_z}{\partial \theta} ,
 \left( \frac{\rho_{pz}}{qR} \frac{\partial \ln T_z}{\partial \theta}  \right)^2
\right\}
\ll 1,
 \end{equation}
 since the strongest poloidal variation of the viscous tensor is dictated by the impurity density.
Additionally, the correspondent viscous \emph{energy} is also actually negligible with respect to the compressional heating in the energy equation,
 \begin{eqnarray}
\frac{\boldsymbol{\pi}_{zdC} : \boldsymbol{\nabla} \mathbf{V}_z}
{p_z \boldsymbol{\nabla} \cdot  \mathbf{V}_z}
\sim
\max \left\{
\frac{\rho_{pz}}{L_{n_z}} \Delta \frac{\lambda_z}{qR},
 \left( \Delta \frac{\lambda_z}{qR} \right)^2 \frac{1}{\Delta} \frac{\partial \ln T_z}{\partial \theta} ,
 \left( \frac{\rho_{pz}}{qR} \frac{\partial \ln T_z}{\partial \theta}  \right)^2
\right\}
\ll 1,
\end{eqnarray}
since
\begin{eqnarray}
\boldsymbol{\pi}_{zdC} : \boldsymbol{\nabla} \mathbf{V}_z
= 
\frac{3}{2}
\left( \frac{\mathbf{B}}{B} \cdot \boldsymbol{\nabla} \mathbf{V}_z \cdot \frac{\mathbf{B}}{B} - \frac{\boldsymbol{\nabla} \cdot \mathbf{V}_z}{3}  \right)
\frac{\mathbf{B}}{B} \cdot \boldsymbol{\pi}_{zdC}  \cdot \frac{\mathbf{B}}{B}.
\end{eqnarray}


\paragraph{Off-diagonal:}
On the one hand, the off-diagonal or gyroviscous ($g$) part of the viscous tensor is obtained on Eq. (44) of \cite{Catto2004}:
\begin{multline}
\boldsymbol{\pi}_{zgC}
= \frac{p_z}{4\Omega_z}
\left\{
\frac{\mathbf{B}}{B} \times 
\left[ 
\left( \boldsymbol{\nabla} \mathbf{V}_z + \frac{2}{5} \frac{\boldsymbol{\nabla} \mathbf{q}_{zC}}{p_z} \right) 
+ \left( \boldsymbol{\nabla} \mathbf{V}_z + \frac{2}{5} \frac{\boldsymbol{\nabla} \mathbf{q}_{zC}}{p_z} \right)^T
\right]
\cdot \left( \mathbf{I} + 3\frac{\mathbf{B}}{B} \frac{\mathbf{B}}{B} \right) 
\right.
\\
\left.
- \left( \mathbf{I} + 3\frac{\mathbf{B}}{B} \frac{\mathbf{B}}{B} \right)
 \cdot
 \left[
 \left( \boldsymbol{\nabla} \mathbf{V}_z + \frac{2}{5} \frac{\boldsymbol{\nabla} \mathbf{q}_{zC}}{p_z} \right) 
+ \left( \boldsymbol{\nabla} \mathbf{V}_z + \frac{2}{5} \frac{\boldsymbol{\nabla} \mathbf{q}_{zC}}{p_z} \right)^T
 \right]
 \times \frac{\mathbf{B}}{B}
 \right\}.
\end{multline}
In order to identify the dominant terms, it is worth bear in mind that the ratio impurity heat flux divided by the pressure to mean flow (\ref{eq:Vz||ord},\ref{eq:Vzperppol},\ref{eq:Vzperprad}) has the following size in each direction
 (\ref{eq:Bxthetapsi}):
\begin{equation}
\begin{aligned}
\frac{\mathbf{q}_{zC}  \cdot \frac{\mathbf{B}}{B}}
{p_z  \mathbf{V}_z \cdot \frac{\mathbf{B}}{B}}
& \sim
\frac{\frac{125 p_z}{32 M_z \nu_{zz}} \frac{\mathbf{B}}{B} \cdot \boldsymbol{\nabla} T_z}
{p_z V_{z\parallel}}
 & \sim
 \frac{\lambda_z}{qR} \frac{\partial \ln T_z}{\partial \theta} \frac{L_{n_z}}{\rho_{pz}},
\\
\frac{ \mathbf{q}_{zC}  \cdot \frac{\mathbf{B} \times \boldsymbol{\nabla} \psi}{B R B_p} }
{p_z \mathbf{V}_z \cdot \frac{\mathbf{B} \times \boldsymbol{\nabla} \psi}{B R B_p} }
& \sim
\frac{ \frac{5 p_z}{2 M_z \Omega_z} \frac{\mathbf{B}}{B} \times \boldsymbol{\nabla} T_z \cdot \frac{\mathbf{B} \times \boldsymbol{\nabla} \psi}{B R B_p}}
{p_z \mathbf{V}_z \cdot \frac{\mathbf{B} \times \boldsymbol{\nabla} \psi}{B R B_p} }
& \sim
\frac{z_i}{z_z} 
\ll 1
\\
\frac{ \mathbf{q}_{zC}  \cdot \frac{\boldsymbol{\nabla} \psi}{R B_p}}
{p_z \mathbf{V}_z \cdot \frac{\boldsymbol{\nabla} \psi}{R B_p}}
& \sim
\frac{ \frac{5 p_z}{2 M_z \Omega_z} \frac{\mathbf{B}}{B} \times \boldsymbol{\nabla} T_z \cdot \frac{\boldsymbol{\nabla} \psi}{R B_p}}
{p_z \mathbf{V}_z \cdot \frac{\boldsymbol{\nabla} \psi}{R B_p} }
&  \sim
\frac{\frac{\partial \ln T_z}{\partial \theta}}{\Delta}
\ll 1.
\end{aligned}
\label{eq:qradpolnegl}
\end{equation}

The size of the gyroviscous force in the \emph{parallel momentum} equation is represented by the following term:
 \begin{eqnarray}
\boldsymbol{\nabla} \cdot \boldsymbol{\pi}_{zgC} \cdot \frac{\mathbf{B}}{B}
\sim
\frac{\boldsymbol{\nabla} p_z}{\Omega_z} \cdot 
\frac{\mathbf{B}}{B} \times 
\left[ 
\left( \boldsymbol{\nabla} \mathbf{V}_z + \frac{2}{5} \frac{\boldsymbol{\nabla} \mathbf{q}_{zC}}{p_z} \right) 
+ \left(  \boldsymbol{\nabla} \mathbf{V}_z + \frac{2}{5} \frac{\boldsymbol{\nabla} \mathbf{q}_{zC}}{p_z} \right)^T
\right]
\cdot \frac{\mathbf{B}}{B},
\label{eq:viscg||momterms}
\end{eqnarray}
where the term with the magnetic field brought inside the gradient is used since the impurity density presents the strongest radial and poloidal variation.
More especifically, the untransposed term in (\ref{eq:viscg||momterms}), 
\begin{eqnarray}
\frac{\boldsymbol{\nabla} p_z}{\Omega_z} \cdot
 \frac{\mathbf{B}}{B} \times 
\left( \boldsymbol{\nabla} V_{z\parallel} + \frac{2}{5} \frac{\boldsymbol{\nabla} q_{z\parallel C}}{p_z} \right) 
\sim
\frac{\Delta p_z}{q R } \frac{\rho_{pz} }{L_{n_z}}  \max \left\{ \frac{\rho_{pz}}{L_{n_z}}, \frac{\lambda_z}{qR}  \frac{\partial \ln T_z}{\partial \theta} \right\} ,
\end{eqnarray}
dominates the transposed one,
\begin{eqnarray}
\left( \nabla_{\parallel}  \mathbf{V}_z + \frac{2}{5} \frac{{\nabla}_{\parallel} \mathbf{q}_{zC}}{p_z} \right)
 \times \frac{\mathbf{B}}{B} 
 \cdot \frac{\boldsymbol{\nabla} p_z}{\Omega_z} 
 \sim
 \frac{\Delta p_z}{qR} \frac{\rho_z^2}{L_{n_z}^2},
\end{eqnarray}
whose size has been estimated by using (\ref{eq:qradpolnegl}) and noticing that
\begin{eqnarray}
\frac{ \mathbf{V}_z \cdot \frac{\boldsymbol{\nabla} \psi}{R B_p} 
\frac{\boldsymbol{\nabla} \psi}{R B_p} \times \frac{\mathbf{B}}{B} 
 \cdot \boldsymbol{\nabla} \ln n_z  }
{ \mathbf{V}_z \cdot \frac{\mathbf{B} \times \boldsymbol{\nabla} \psi}{B R B_p}\frac{\boldsymbol{\nabla} \psi}{R B_p} \cdot \boldsymbol{\nabla} \ln n_z }
\sim
\frac{ \Delta \frac{\rho_{pz}}{qR} v_{Tz} 
\frac{B }{qR B_p} 
\frac{\partial \ln n_z}{\partial \theta} }
{ \frac{\rho_z}{L_{n_z}} v_{Tz}  R B_p \frac{\partial \ln n_z}{\partial \psi} }
\sim
\left( \Delta \frac{L_{n_z}}{\epsilon R} \right)^2
\ll 1.
\label{eq:VzcrossBdotgrad}
\end{eqnarray}
In conclusion, the gyroviscous force can indeed be successfully neglected in the parallel momentum equation:
 \begin{eqnarray}
 \frac{\boldsymbol{\nabla} \cdot \boldsymbol{\pi}_{zgC} \cdot \frac{\mathbf{B}}{B}}
 {{\nabla}_{\parallel} p_z + z_z e n_z {\nabla}_{\parallel} \Phi }
 \sim
 \max \left\{
\frac{\rho^2_{pz}}{L^2_{n_z}}, \frac{\rho_{pz} }{L_{n_z}} \frac{\lambda_z}{qR}  \frac{\partial \ln T_z}{\partial \theta}
 \right\}
 \ll 1.
\end{eqnarray}

The size of the gyroviscous \emph{energy} can be estimated by using the following terms since the strongest variation of the impurity flow is given by the impurity density:
\begin{multline}
\boldsymbol{\pi}_{zgC} : \boldsymbol{\nabla} \mathbf{V}_z
\sim
\mathbf{V}_{z\perp} \cdot 
\frac{\mathbf{B}}{B} \times 
\left[ 
\left( \boldsymbol{\nabla} \mathbf{V}_z + \frac{2}{5} \frac{\boldsymbol{\nabla} \mathbf{q}_{zC}}{p_z} \right) 
+ \left( \boldsymbol{\nabla} \mathbf{V}_z \right)^T
\right]
\cdot \frac{\boldsymbol{\nabla} p_z}{4\Omega_z} 
 \\
- \mathbf{V}_z \cdot 
 \left[
\boldsymbol{\nabla} \mathbf{V}_z
+ \left( \boldsymbol{\nabla} \mathbf{V}_z + \frac{2}{5} \frac{\boldsymbol{\nabla} \mathbf{q}_{zC}}{p_z} \right)^T
 \right]
 \times \frac{\mathbf{B}}{B}
\cdot \frac{\boldsymbol{\nabla} p_z}{4\Omega_z} .
\label{eq:viscenergy}
\end{multline}
Here it has been used that the perpendicular mean flow is larger than the perpendicular heat flux (\ref{eq:qradpolnegl}).
The size of each term in (\ref{eq:viscenergy}) is calculated by using also (\ref{eq:Vzgradord}) and (\ref{eq:VzcrossBdotgrad}):
\begin{equation}
\begin{aligned}
\mathbf{V}_{z\perp} \cdot
\frac{\mathbf{B}}{B} \times 
\left( \boldsymbol{\nabla} \mathbf{V}_z + \frac{2}{5} \frac{\boldsymbol{\nabla} \mathbf{q}_{zC}}{p_z} \right) 
\cdot \frac{\boldsymbol{\nabla} p_z}{4\Omega_z}
& \sim
 \frac{\Delta p_z v_{Tz}}{qR}  
\frac{\rho_z^2}{L_{n_z}^2}
\max \left\{ \frac{\rho_{pz}}{L_{n_z}}, \frac{\lambda_z}{qR} \frac{\partial \ln T_z}{\partial \theta} \right\} ,
\\
\frac{\boldsymbol{\nabla} p_z }{4 \Omega_z} \cdot 
 \boldsymbol{\nabla} \mathbf{V}_{z\perp}
\times \frac{\mathbf{B}}{B} \cdot \mathbf{V}_{z\perp}
& \sim
\frac{\Delta p_z v_{Tz}}{ qR} 
\frac{\rho_z^2}{L_{n_z}^2}  \frac{\rho_{pz}}{L_{n_z}},
\\
\mathbf{V}_z \cdot \boldsymbol{\nabla} \mathbf{V}_z 
 \times \frac{\mathbf{B}}{B}
\cdot \frac{\boldsymbol{\nabla} p_z}{4 \Omega_z}
& \sim
\frac{\Delta p_z v_{Tz}}{qR} 
\frac{\rho_z^2}{L_{n_z}^2} \frac{\rho_{pz}}{L_{n_z}}
\\
\frac{\boldsymbol{\nabla} p_z}{4 \Omega_z} \cdot 
\frac{\mathbf{B}}{B} \times 
 \left( \boldsymbol{\nabla} \mathbf{V}_z + \frac{2}{5} \frac{\boldsymbol{\nabla} \mathbf{q}_{zC}}{ p_z } \right)
\cdot \mathbf{V}_z 
& \sim
\frac{\Delta p_z  v_{Tz}}{qR }  \frac{\rho_{pz}^2}{L_{n_z}^2} 
 \max \left\{ \frac{\rho_{pz}}{L_{n_z}}, \frac{\lambda_z}{qR} \frac{\partial \ln T_z}{\partial \theta} \right\} .
\end{aligned}
\end{equation}
The viscous energy can hence be neglected in the energy conservation equation since
\begin{eqnarray}
\frac{\boldsymbol{\pi}_{zgC} : \boldsymbol{\nabla} \mathbf{V}_z}
{p_z \boldsymbol{\nabla} \cdot \mathbf{V}_z}
\sim
 \max \left\{ \frac{\rho_{pz}^2}{L_{n_z}^2}, 
 \frac{\rho_{pz}}{L_{n_z}} \frac{\lambda_z}{qR} \frac{\partial \ln T_z}{\partial \theta} \right\} 
 \ll 1.
\end{eqnarray}
\section{Checking assumptions of the derivation of the velocity\label{sec:checking_a}}

The perpendicular impurity flow,
\begin{eqnarray}
\mathbf{V}_{z\perp} 
= \frac{c}{B^2} \mathbf{B} \times \left( \boldsymbol{\nabla} \Phi + \frac{\boldsymbol{\nabla}  p_z + M_z n_z \mathbf{V}_z \cdot \boldsymbol{\nabla} \mathbf{V}_z + \boldsymbol{\nabla} \cdot \boldsymbol{\pi}_z + \mathbf{R}_{iz} }{z_z e n_z} \right),
\label{eq:Vzperptotal}
\end{eqnarray}
 is simplified into (\ref{eq:Vzperp}) by assuming that the perpendicular projection of the inertia, friction and viscous force are negligible. 
The sizes of these terms are estimated \emph{a posteriori} in this appendix by using the resulting impurity flow (\ref{eq:Vz}) to check the applicability of the approximations.

\subsection{Neglected inertial term} 
The inertial term can be conveniently rewritten as a divergence, by using conservation of impurity particles (\ref{eq:conspartparperp}), to find
\begin{eqnarray}
n_z \mathbf{V}_z \cdot \boldsymbol{\nabla} \mathbf{V}_z
 =
 \boldsymbol{\nabla} \cdot \left( n_z \mathbf{V}_z \mathbf{V}_z \right).
\end{eqnarray}
The projection of the inertial contribution to the perpendicular impurity flow (\ref{eq:Vzperptotal})  in the directions perpendicular to the flux surface and within the latter but perpendicular to the magnetic field are then respectively given by:
\begin{equation}
\frac{\mathbf{B}}{B}  \times \frac{ \boldsymbol{\nabla} \cdot \left( n_z \mathbf{V}_z \mathbf{V}_z \right)}{\Omega_z n_z}  \cdot \frac{\boldsymbol{\nabla} \psi}{R B_p}
 = -  \frac{ \boldsymbol{\nabla} \cdot \left( n_z \mathbf{V}_z \mathbf{V}_z \cdot \frac{\mathbf{B} \times \boldsymbol{\nabla} \psi}{B R B_p}  \right)}{\Omega_z n_z}  
+ \frac{\mathbf{V}_z}{\Omega_z}  \cdot \boldsymbol{\nabla} \left(\frac{\mathbf{B} \times \boldsymbol{\nabla} \psi}{B R B_p} \right) \cdot \mathbf{V}_z 
\label{eq:inertiaprojrad}
\end{equation}
and
\begin{equation}
\frac{\mathbf{B}}{B}  \times \frac{ \boldsymbol{\nabla} \cdot \left( n_z \mathbf{V}_z \mathbf{V}_z \right)}{\Omega_z n_z}  \cdot \frac{\mathbf{B} \times \boldsymbol{\nabla} \psi}{B R B_p}
 =  \frac{ \boldsymbol{\nabla} \cdot \left( n_z \mathbf{V}_z \mathbf{V}_z \cdot \frac{\boldsymbol{\nabla} \psi}{R B_p} \right)}{\Omega_z n_z}  
- \frac{1}{\Omega_z} \mathbf{V}_z \cdot \boldsymbol{\nabla} \left( \frac{\boldsymbol{\nabla} \psi}{R B_p} \right) \cdot \mathbf{V}_z .
\label{eq:inertiaprojpol}
\end{equation}

Since the impurity density presents the strongest poloidal (\ref{eq:ddthetaorderings}) and radial (\ref{eq:ddpsiorderings}) variation, the first terms on the right hand of (\ref{eq:inertiaprojrad}) and (\ref{eq:inertiaprojpol}) are used to determine the size of the correspondent projection of the inertial contribution to the perpendicular impurity flow (\ref{eq:divperpflux},\ref{eq:Vzperprad},\ref{eq:Vzperppol}):
\begin{equation}
\begin{aligned}
\frac{
\boldsymbol{\nabla} \cdot \left( n_z \mathbf{V}_z \mathbf{V}_z \cdot \frac{\mathbf{B} \times \boldsymbol{\nabla} \psi}{B R B_p} \right)
}{\Omega_z n_z}  
= &
\frac{\mathbf{B} \cdot \boldsymbol{\nabla} \theta}{\Omega_z n_z}  
 \frac{\partial}{\partial \psi} \left( \frac{n_z \mathbf{V}_z \cdot \boldsymbol{\nabla} \psi \mathbf{V}_z \cdot \frac{\mathbf{B} \times \boldsymbol{\nabla} \psi}{B R B_p}}{\mathbf{B} \cdot \boldsymbol{\nabla} \theta} \right)
& \sim
 \Delta \frac{\rho_{pz}}{qR} v_{Tz}
\frac{\rho_z^2}{L_{n_z}^2} 
 \\
& +  
\frac{\mathbf{B} \cdot \boldsymbol{\nabla} \theta}{\Omega_z n_z}  
\frac{\partial}{\partial \theta} \left( \frac{n_z \mathbf{V}_z \cdot \boldsymbol{\nabla} \theta \mathbf{V}_z \cdot \frac{\mathbf{B} \times \boldsymbol{\nabla} \psi}{B R B_p}}{\mathbf{B} \cdot \boldsymbol{\nabla} \theta} \right)
& \sim
 \Delta \frac{\rho_{pz}}{qR} v_{Tz}
\frac{\rho_z^2}{L_{n_z}^2} 
\end{aligned}
\end{equation}
and
\begin{equation}
\begin{aligned}
\frac{
\boldsymbol{\nabla} \cdot \left( n_z \mathbf{V}_z \mathbf{V}_z \cdot \frac{\boldsymbol{\nabla} \psi}{R B_p} \right)
}{\Omega_z n_z}  
= &
\frac{\mathbf{B} \cdot \boldsymbol{\nabla} \theta}{\Omega_z n_z}  
 \frac{\partial}{\partial \psi} \left( \frac{n_z \mathbf{V}_z \cdot \boldsymbol{\nabla} \psi \mathbf{V}_z \cdot \frac{\boldsymbol{\nabla} \psi}{R B_p}}{\mathbf{B} \cdot \boldsymbol{\nabla} \theta} \right)
& \sim
 \frac{\rho_z}{L_{n_z}} v_{Tz} \left( \Delta \frac{\rho_{pz}}{qR} \right)^2
 \\
& +  
\frac{\mathbf{B} \cdot \boldsymbol{\nabla} \theta}{\Omega_z n_z}  
\frac{\partial}{\partial \theta} \left( \frac{n_z \mathbf{V}_z \cdot \boldsymbol{\nabla} \theta \mathbf{V}_z \cdot \frac{\boldsymbol{\nabla} \psi}{R B_p}}{\mathbf{B} \cdot \boldsymbol{\nabla} \theta} \right)
\hspace{-0.5cm}
& \sim
 \frac{\rho_z}{L_{n_z}} v_{Tz} \left( \Delta \frac{\rho_{pz}}{qR} \right)^2.
\end{aligned}
\end{equation}
In conclusion, the projection of the inertial contribution to the perpendicular impurity flow (\ref{eq:Vzperptotal}) can be successfully neglected
since
\begin{equation}
\frac{
\frac{\mathbf{B}}{B}  \times \frac{ \boldsymbol{\nabla} \cdot \left( n_z \mathbf{V}_z \mathbf{V}_z \right)}{\Omega_z n_z}  \cdot \frac{\boldsymbol{\nabla} \psi}{R B_p}
}{
\mathbf{V}_{z\perp} \cdot \frac{\boldsymbol{\nabla} \psi}{R B_p}
}
\sim
\frac{
\frac{
\boldsymbol{\nabla} \cdot \left( n_z \mathbf{V}_z \mathbf{V}_z \cdot \frac{\mathbf{B} \times \boldsymbol{\nabla} \psi}{B R B_p} \right)
}{\Omega_z n_z}  
}{
\mathbf{V}_{z\perp} \cdot \frac{\boldsymbol{\nabla} \psi}{R B_p}
}
\sim
\frac{\rho_z^2}{L_{n_z}^2} 
\ll 1
\label{eq:inertiaradnegl}
\end{equation}
and
\begin{equation}
\frac{
\frac{\mathbf{B}}{B}  \times \frac{ \boldsymbol{\nabla} \cdot \left( n_z \mathbf{V}_z \mathbf{V}_z \right)}{\Omega_z n_z}  \cdot \frac{\mathbf{B} \times \boldsymbol{\nabla} \psi}{B R B_p}
}{
\mathbf{V}_{z\perp} \cdot \frac{\mathbf{B} \times \boldsymbol{\nabla} \psi}{B R B_p}
}
\sim
\frac{
\frac{
\boldsymbol{\nabla} \cdot \left( n_z \mathbf{V}_z \mathbf{V}_z \cdot \frac{\boldsymbol{\nabla} \psi}{R B_p} \right)
}{\Omega_z n_z}    
}{
\mathbf{V}_{z\perp} \cdot \frac{\mathbf{B} \times \boldsymbol{\nabla} \psi}{B R B_p}
}
\sim
 \left( \Delta \frac{\rho_{pz}}{qR} \right)^2
\ll 1.
\end{equation}
Inertial terms have zero flux surface average regardless of ordering.

\subsection{Neglected friction term} 

The size of the components of the perpendicular friction in the direction perpendicular to the flux surface and perpendicular to the magnetic field but contained within the flux function are respectfully given by 
\begin{equation}
\mathbf{R}_{iz} \cdot \frac{\boldsymbol{\nabla} \psi}{R B_p}
 =
M_i \int d^3 v_i \mathbf{w}_i \cdot \frac{\boldsymbol{\nabla} \psi}{R B_p} \left( C_{iz1} - \left< C_{iz1} \right>_{\varphi} \right)
 \sim
M_i n_i \nu_{iz} v_{Tz} \Delta \frac{\rho_{pz}}{qR}
\end{equation}
and
\begin{equation}
\mathbf{R}_{iz} \cdot \frac{\mathbf{B} \times \boldsymbol{\nabla} \psi}{B R B_p}
 =
M_i \int d^3 v_i \mathbf{w}_i \cdot \frac{\mathbf{B} \times \boldsymbol{\nabla} \psi}{B R B_p} \left( C_{iz1} - \left< C_{iz1} \right>_{\varphi} \right)
 \sim
M_i n_i  \nu_{iz} v_{Tz} \frac{\rho_z}{L_{n_z}};
\label{eq:Rizpolord}
\end{equation}
since the gyrophase-dependent piece of the unlike collision operator (\ref{eq:Ciz1fsa}) is
\begin{equation}
C_{iz1} - \left< C_{iz1} \right>_{\varphi}
= \frac{3 \sqrt{2 \pi} \nu_{iz} T_i^{\frac{3}{2}}}{4 M_i^{\frac{3}{2}}}
\left[
\boldsymbol{\nabla}_{v_i} \cdot 
\left(
\boldsymbol{\nabla}_{v_i}  \boldsymbol{\nabla}_{v_i}  v_i
\cdot
\boldsymbol{\nabla}_{v_i} \tilde{f}_{i1} 
\right)
+ 2  f_{iM\left< \right>} \frac{M_i}{T_i} \frac{\mathbf{v}_{i\perp}}{v_i^3}
\cdot \mathbf{V}_z
\right],
\label{eq:Ciz1gd}
\end{equation}
with $\tilde{f}_{i1} = f_{i1} - \bar{f}_{i1}$
and the size of the perpendicular impurity flow dictated by (\ref{eq:Vzperprad}) and (\ref{eq:Vzperppol}).
In conclusion, the contribution of the friction force to the perpendicular impurity flow (\ref{eq:Vzperptotal}) can be successfully neglected if
\begin{eqnarray}
\frac{
\frac{\mathbf{B}}{B} \times \frac{\mathbf{R}_{iz}}{M_z \Omega_z n_z} \cdot \frac{\boldsymbol{\nabla} \psi}{R B_p}
}{
\mathbf{V}_{z\perp} \cdot \frac{\boldsymbol{\nabla} \psi}{R B_p}
}
=
\frac{
- \frac{\mathbf{R}_{iz}}{M_z \Omega_z n_z} \cdot \frac{\mathbf{B} \times \boldsymbol{\nabla} \psi}{B R B_p} 
}{
\mathbf{V}_{z\perp} \cdot \frac{\boldsymbol{\nabla} \psi}{R B_p}
}
\sim
\frac{1}{\Delta} \frac{qR}{\lambda_z} \frac{\rho_{pz}}{L_{n_z}} \sqrt{\frac{z_z}{z_i}}  \frac{B_p^2}{B^2} 
\ll 1,
\label{eq:perpmomfricrad}
\end{eqnarray}
which implies that (\ref{eq:Vzradialsmall})
\begin{eqnarray}
\frac{
\frac{\mathbf{B}}{B}  \times \frac{\mathbf{R}_{iz}}{M_z \Omega_z n_z}  \cdot \frac{\mathbf{B} \times \boldsymbol{\nabla} \psi}{B R B_p}
}{
\mathbf{V}_{z\perp} \cdot \frac{\mathbf{B} \times \boldsymbol{\nabla} \psi}{B R B_p}
}
=
\frac{
\frac{\mathbf{R}_{iz}}{M_z \Omega_z n_z} \cdot \frac{\boldsymbol{\nabla} \psi}{R B_p} 
}{
\mathbf{V}_{z\perp} \cdot \frac{\mathbf{B} \times \boldsymbol{\nabla} \psi}{B R B_p}
}
\sim
\frac{1}{\Delta} \frac{qR}{\lambda_z} \frac{\rho_{pz}}{L_{n_z}}  \sqrt{\frac{z_z}{z_i}} \left( \Delta \frac{L_{n_z}}{qR} \right)^2
\hspace{-0.25cm}
\ll 1.
\label{eq:perpmomfricpol}
\end{eqnarray}

\subsection{Neglected viscous term} 

\paragraph{Diagonal terms:}
The divergence of the viscous tensor dotted into the perpendicular directional vector can be related (\ref{eq:Bxthetapsi}) to that of the parallel diagonal component, whose size is estimated on (\ref{eq:pizdb}), to find
 \begin{equation}
\boldsymbol{\nabla} \cdot \left( \boldsymbol{\pi}_{zdC} \cdot \frac{\boldsymbol{\nabla} \psi}{R B_p} \right)
 = 
- \frac{1}{2} \boldsymbol{\nabla} \cdot \left( \frac{\boldsymbol{\nabla} \psi}{R B_p}
\frac{\mathbf{B}}{B} \cdot \boldsymbol{\pi}_{zdC} \cdot \frac{\mathbf{B}}{B}  \right)
 \sim
\frac{1}{L_{n_z}} \frac{\mathbf{B}}{B} \cdot \boldsymbol{\pi}_{zdC} \cdot \frac{\mathbf{B}}{B}
\end{equation}
and
\begin{equation}
\boldsymbol{\nabla} \cdot \left( \boldsymbol{\pi}_{zdC} \cdot \frac{\mathbf{B} \times \boldsymbol{\nabla} \psi}{B R B_p}  \right)
 = 
- \frac{1}{2} \boldsymbol{\nabla} \cdot \left( \frac{\mathbf{B} \times \boldsymbol{\nabla} \psi}{B R B_p}
\frac{\mathbf{B}}{B} \cdot \boldsymbol{\pi}_{zdC} \cdot \frac{\mathbf{B}}{B}  \right)
 \sim
\frac{\Delta B}{q R B_p} \frac{\mathbf{B}}{B} \cdot  \boldsymbol{\pi}_{zdC} \cdot \frac{\mathbf{B}}{B}.
\end{equation}
Consequently, the contribution of the diagonal terms of the viscous tensor to the perpendicular impurity flow (\ref{eq:Vzperptotal}) can be automatically neglected with no further assumptions since
\begin{multline}
\frac{
\frac{\mathbf{B}}{B} \times \frac{\boldsymbol{\nabla} \cdot \boldsymbol{\pi}_{zdC}}{M_z \Omega_z n_z} \cdot \frac{\boldsymbol{\nabla} \psi}{R B_p}
}{
\mathbf{V}_{z\perp} \cdot \frac{\boldsymbol{\nabla} \psi}{R B_p}
}
=
\frac{
- \frac{\boldsymbol{\nabla} \cdot \boldsymbol{\pi}_{zdC}}{M_z \Omega_z n_z} \cdot \frac{\mathbf{B} \times \boldsymbol{\nabla} \psi}{B R B_p} 
}{
\mathbf{V}_{z\perp} \cdot \frac{\boldsymbol{\nabla} \psi}{R B_p}
}
\sim
\frac{
\frac{\mathbf{B}}{B}  \times \frac{\boldsymbol{\nabla} \cdot \boldsymbol{\pi}_{zdC}}{M_z \Omega_z n_z}  \cdot \frac{\mathbf{B} \times \boldsymbol{\nabla} \psi}{B R B_p}
}{
\mathbf{V}_{z\perp} \cdot \frac{\mathbf{B} \times \boldsymbol{\nabla} \psi}{B R B_p}
}
=
\frac{
\frac{\boldsymbol{\nabla} \cdot \boldsymbol{\pi}_{zdC}}{M_z \Omega_z n_z} \cdot \frac{\boldsymbol{\nabla} \psi}{R B_p} 
}{
\mathbf{V}_{z\perp} \cdot \frac{\mathbf{B} \times \boldsymbol{\nabla} \psi}{B R B_p}
}
\sim
 \\
\sim
\max \left\{
\Delta  \frac{\rho_{pz}}{L_{n_z}}  \frac{\lambda_z}{ qR},
\Delta \left(\frac{\lambda_z}{qR} \right)^2\frac{\partial \ln T_z}{\partial \theta} ,
\left( \frac{\rho_{pz}}{qR}  \frac{\partial \ln T_z}{\partial \theta}  \right)^2
\right\}
\ll 1.
\end{multline}
 
\paragraph{Off-diagonal terms:}
 
The size of the divergence of the dominant terms (\ref{eq:qradpolnegl}) of the gyroviscous tensor dotted into the unitary vector perpendicular to the flux surface,
\begin{multline}
\boldsymbol{\pi}_{zgC} \cdot \frac{\boldsymbol{\nabla} \psi}{R B_p}
=
\frac{p_z}{4 \Omega_z}
\left\{
 \frac{\mathbf{B}}{B} \times 
\left[  \boldsymbol{\nabla} \mathbf{V}_z 
+ \left( \boldsymbol{\nabla} \mathbf{V}_z \right)^T \right]
\cdot \frac{\boldsymbol{\nabla} \psi}{R B_p}
\right.
 \\ 
\left.
- \left( \mathbf{I} + 3\frac{\mathbf{B}}{B} \frac{\mathbf{B}}{B} \right)
 \cdot
 \left[
 \boldsymbol{\nabla} \mathbf{V}_z
+ \left( \boldsymbol{\nabla} \mathbf{V}_z + \frac{2}{5} \frac{ \boldsymbol{\nabla} \mathbf{q}_{zC} }{p_z} \right)^T
 \right]
\cdot \frac{ \mathbf{B} \times \boldsymbol{\nabla} \psi}{B R B_p}
\right\} ,
\end{multline}
is estimated by applying the gradient to the impurity pressure
and analysing each term separately:
\begin{equation}
\begin{aligned}
\frac{\boldsymbol{\nabla} p_z}{4 \Omega_z} \cdot \frac{\mathbf{B}}{B} \times \boldsymbol{\nabla}  \mathbf{V}_z \cdot  \frac{\boldsymbol{\nabla} \psi}{R B_p}
& \sim
 \frac{\Delta^2 p_z}{qR} \frac{\rho_{pz}}{qR}  \frac{\rho_{pz}}{L_{nz}} ,
 \\
\frac{\boldsymbol{\nabla} \psi}{R B_p} \cdot
 \boldsymbol{\nabla} \mathbf{V}_z  \times  \frac{\mathbf{B}}{B} \cdot \frac{\boldsymbol{\nabla} p_z}{4 \Omega_z} 
& \sim
\frac{p_z}{L_{n_z}} \frac{\rho^2_z}{L^2_{n_z}},
\\
\frac{\boldsymbol{\nabla} p_z}{4 \Omega_z} \cdot
 \boldsymbol{\nabla} \mathbf{V}_z
\cdot \frac{ \mathbf{B} \times \boldsymbol{\nabla} \psi}{B R B_p}
& \sim
\frac{p_z}{L_{n_z}} \frac{\rho_z^2}{L_{n_z}^2}
 \\
\frac{ \mathbf{B} \times \boldsymbol{\nabla} \psi}{B R B_p} 
\cdot \left( \boldsymbol{\nabla} \mathbf{V}_z + \frac{2}{5} \frac{\boldsymbol{\nabla} \mathbf{q}_{zC}}{p_z} \right)
\cdot  \frac{\boldsymbol{\nabla} p_z}{4 \Omega_z}
& \sim
\frac{\Delta^2 p_z}{q R} \frac{\rho_{pz}}{qR}
\max \left\{
\frac{\rho_{pz}}{L_{n_z}},
\frac{\lambda_z}{qR} \frac{\partial \ln T_z}{\partial \theta}
\right\}.
\end{aligned}
\end{equation}

Analogously, the size of the divergence of the dominant terms (\ref{eq:qradpolnegl}) of the gyroviscous tensor dotted into the unitary vector within the flux surface but perpendicular to the magnetic field, 
 \begin{multline}
\boldsymbol{\pi}_{zgC} \cdot \frac{\mathbf{B} \times \boldsymbol{\nabla} \psi}{B R B_p}
=
\frac{p_z}{4 \Omega_z}
\left\{
 \frac{\mathbf{B}}{B} \times 
\left[ 
 \boldsymbol{\nabla} \mathbf{V}_z
+ \left(\boldsymbol{\nabla} \mathbf{V}_z \right)^T
\right]
\cdot \frac{\mathbf{B} \times \boldsymbol{\nabla} \psi}{B R B_p}
\right.
 \\
\left.
+ \left( \mathbf{I} + 3\frac{\mathbf{B}}{B} \frac{\mathbf{B}}{B} \right)
 \cdot
 \left[
 \boldsymbol{\nabla} \mathbf{V}_z
+ \left( \boldsymbol{\nabla} \mathbf{V}_z + \frac{2}{5} \frac{\boldsymbol{\nabla} \mathbf{q}_{zC}}{p_z} \right)^T
 \right]
 \cdot \frac{\boldsymbol{\nabla} \psi}{R B_p}
\right\} ,
\end{multline}
is estimated term by term as follows:
\begin{equation}
\begin{aligned}
\frac{\boldsymbol{\nabla} p_z }{4 \Omega_z} \cdot
\frac{\mathbf{B}}{B} \times 
 \boldsymbol{\nabla} \mathbf{V}_z
\cdot \frac{\mathbf{B} \times \boldsymbol{\nabla} \psi}{B R B_p}
& \sim
\frac{\Delta p_z}{qR} \frac{\rho_{pz}}{L_{nz}} \frac{\rho_z}{L_{n_z}},
\\
 \frac{\mathbf{B} \times \boldsymbol{\nabla} \psi}{B R B_p} \cdot 
\boldsymbol{\nabla} \mathbf{V}_z
 \times \frac{\mathbf{B}}{B} \cdot \frac{\boldsymbol{\nabla}  p_z}{4 \Omega_z} 
&  \sim
 \frac{\Delta p_z}{q R}  \frac{\rho_{pz}}{L_{n_z}} \frac{\rho_z}{L_{n_z}},
 \\
 \frac{\boldsymbol{\nabla} p_z}{4 \Omega_z} \cdot
 \boldsymbol{\nabla} \mathbf{V}_z
 \cdot \frac{\boldsymbol{\nabla} \psi}{R B_p}
& \sim
  \frac{\Delta p_z}{qR} \frac{\rho_{pz}}{L_{nz}}  \frac{\rho_z}{L_{nz}}
  \\
\frac{\boldsymbol{\nabla} \psi}{R B_p} \cdot
 \left( p_z \boldsymbol{\nabla} \mathbf{V}_z + \frac{2}{5} \boldsymbol{\nabla} \mathbf{q}_{zC} \right)
 \cdot \frac{\boldsymbol{\nabla} \ln n_z}{4 \Omega_z} 
& \sim
\frac{\Delta p_z}{qR} \frac{\rho_z}{L_{n_z}}
\max \left\{
 \frac{\rho_{pz}}{L_{n_z}},
 \frac{\lambda_z}{qR} \frac{\partial \ln T_z}{\partial \theta} 
\right\}.
\end{aligned}
\end{equation}

The contribution of the gyroviscous tensor to the perpendicular impurity flow (\ref{eq:Vzperptotal}) can hence be neglected as well since
\begin{equation}
\frac{
\frac{\mathbf{B}}{B} \times \frac{\boldsymbol{\nabla} \cdot \boldsymbol{\pi}_{zgC}}{M_z \Omega_z n_z} \cdot \frac{\boldsymbol{\nabla} \psi}{R B_p}
}{
\mathbf{V}_{z\perp} \cdot \frac{\boldsymbol{\nabla} \psi}{R B_p}
}
=
\frac{
- \frac{\boldsymbol{\nabla} \cdot \boldsymbol{\pi}_{zgC}}{M_z \Omega_z n_z} \cdot \frac{\mathbf{B} \times \boldsymbol{\nabla} \psi}{B R B_p} 
}{
\mathbf{V}_{z\perp} \cdot \frac{\boldsymbol{\nabla} \psi}{R B_p}
}
\sim
\frac{B_p}{B} \frac{\rho_z}{L_{nz}}
\max \left\{
\frac{\rho_{pz}}{L_{nz}},
 \frac{\lambda_z}{qR} \frac{\partial \ln T_z}{\partial \theta} 
\right\}
\ll 1
\label{eq:viscradnegl}
\end{equation}
and
\begin{equation}
\frac{
\frac{\mathbf{B}}{B}  \times \frac{\boldsymbol{\nabla} \cdot \boldsymbol{\pi}_{zgC}}{M_z \Omega_z n_z}  \cdot \frac{\mathbf{B} \times \boldsymbol{\nabla} \psi}{B R B_p}
}{
\mathbf{V}_{z\perp} \cdot \frac{\mathbf{B} \times \boldsymbol{\nabla} \psi}{B R B_p}
}
=
\frac{
\frac{\boldsymbol{\nabla} \cdot \boldsymbol{\pi}_{zgC}}{M_z \Omega_z n_z} \cdot \frac{\boldsymbol{\nabla} \psi}{R B_p} 
}{
\mathbf{V}_{z\perp} \cdot \frac{\mathbf{B} \times \boldsymbol{\nabla} \psi}{B R B_p}
}
\sim
\max \left\{
 \frac{\rho^2_z}{L^2_{n_z}}
, \Delta^2 \frac{L_{n_z}}{qR} \frac{\rho_{pz}}{qR} \frac{\lambda_z}{qR} \frac{\partial \ln T_z}{\partial \theta}
\right\}
\ll 1.
\end{equation}

\subsection{Parallel impurity flow calculation} 

If the perpendicular projection of the inertia, friction and divergence of the anisotropic pressure tensor are  negligible on the perpendicular momentum conservation (\ref{eq:Vzperp}), their correspondent contribution to the conservation of particles (\ref{eq:divperpflux}) are also negligible under the stablished orderings.

\section{Energy conservation equation \label{sec:temperatureeq}}

The impurity energy equation balances convection, compressional heating, viscous energy, the divergence of the conductive heat flux, $\mathbf{q}_z$ and equilibration,
\begin{eqnarray}
\frac{3}{2} n_z \mathbf{V}_z \cdot \boldsymbol{\nabla} T_z 
+ p_z \boldsymbol{\nabla} \cdot \mathbf{V}_z
 + \boldsymbol{\pi}_z : \boldsymbol{\nabla} \mathbf{V}_z  
 + \boldsymbol{\nabla} \cdot \mathbf{q}_z 
 =
 \frac{3}{2}  \nu_{zi} n_z \left( T_i - T_z \right).
 \label{eq:energyeqchap7}
 \end{eqnarray}
 
Note that the compressional heating in \ref{eq:energyeqchap7} can be rewritten by using conservation of impurity particles as 
\begin{eqnarray}
p_z \boldsymbol{\nabla} \cdot \mathbf{V}_z
=
- T_z \mathbf{V}_z \cdot \boldsymbol{\nabla} n_z .
\end{eqnarray}
Importantly, even when the radial component of the impurity flow (\ref{eq:Vzperprad}) is very small with respect to the poloidal impurity flow (\ref{eq:Vzperppol},\ref{eq:Vz||ord}), 
\begin{eqnarray}   
\frac{\mathbf{V}_{z\perp} \cdot \frac{\boldsymbol{\nabla} \psi}{R B_p}}{V_{z\parallel} \frac{B_p}{B}}
\sim
\frac{\mathbf{V}_{z\perp} \cdot \frac{\boldsymbol{\nabla} \psi}{R B_p}}{\mathbf{V}_{z\perp} \cdot \frac{\mathbf{B} \times \boldsymbol{\nabla} \psi}{B R B_p}}
\sim
\Delta \frac{L_{nz}}{\epsilon R}
\ll 1,
\label{eq:Vzradialsmall}
\end{eqnarray}
the radial variation of the impurity density is strong enough, recall (\ref{eq:ddthetaorderings}), that its divergence can compete with the divergence of the poloidal flow. 
This implies that both components of the perpendicular impurity flow must  be retained in the conservation equations when the impurity flow is dotted into a gradient of the impurity density
since
\begin{eqnarray}
\mathbf{V}_z \cdot \boldsymbol{\nabla} \ln n_z
\sim
\mathbf{V}_{z\perp} \cdot \boldsymbol{\nabla} \psi \frac{\partial \ln n_z}{\partial \psi}
\sim
V_z \cdot \boldsymbol{\nabla} \theta \frac{\partial \ln n_z}{\partial \theta}
\sim
\Delta \frac{\rho_{pz}}{L_{nz}} \frac{v_{Tz}}{qR}.
\label{eq:Vzgradord}
\end{eqnarray}

When the friction is taken to compete with the pressure and potential gradient terms in parallel momentum conservation, 
then the equilibration term dominates over the compressional heating term in the energy equation since
\begin{eqnarray}
\frac{
\frac{p_z \boldsymbol{\nabla} \cdot \mathbf{V}_z}
{\frac{3}{2} n_z \nu_{zi} \left( T_i - T_z \right)} 
}{
\frac{\frac{\mathbf{B}}{B} \cdot \boldsymbol{\nabla} p_z}{R_{zi\parallel}} 
}
\sim
\frac{\Delta \frac{ \lambda_z}{qR} \sqrt{\frac{z_z}{z_i}} \frac{\rho_{pz}}{L_{n_z}}}
{\Delta \frac{ \lambda_z}{qR} \sqrt{\frac{z_z}{z_i}} \frac{L_{n_z}}{\rho_{pz}}}
\sim
\left( \frac{{\rho}_{pz}}{L_{n_z}} \right)^2
\ll 1.
\end{eqnarray}
Note that this implies that viscous effects can also be neglected in the energy equation:
\begin{equation}
\frac{\boldsymbol{\pi}_{zdC} : \boldsymbol{\nabla} \mathbf{V}_z}
{p_z \boldsymbol{\nabla} \cdot \mathbf{V}_z }
 \sim
\max \left\{
\frac{\rho_{pz}}{L_{n_z}} \Delta \frac{\lambda_z}{ qR},
\Delta \frac{\lambda_z^2}{q^2R^2} \frac{\partial \ln T_z}{\partial \theta} ,
\left( \frac{\rho_{pz}}{qR}  \frac{\partial \ln T_z}{\partial \theta}  \right)^2
\right\}
\ll 1,
\label{eq:neglviscdenergy}
\end{equation}
\begin{equation}
\frac{ \boldsymbol{\pi}_{zgC} : \boldsymbol{\nabla} \mathbf{V}_z }
{p_z \boldsymbol{\nabla} \cdot \mathbf{V}_z }
 \sim
 \max \left\{ 
 \frac{\rho_{pz}^2}{L_{n_z}^2}, 
 \frac{\rho_{pz}}{L_{n_z}} \frac{\lambda_z}{qR} \frac{\partial \ln T_z}{\partial \theta}
  \right\} 
 \ll 1
 \label{eq:neglviscodenergy}
\end{equation}
Finally, the divergence of the heat flux can be neglected when:
\begin{equation}
\begin{aligned}
\frac{ \boldsymbol{\nabla} \cdot \mathbf{q}_{zC}}
{\frac{3}{2} n_z \nu_{zi} \left( T_i - T_z \right)}
& \sim  
\max \left\{
\frac{z_i}{z_z}, 
\frac{1}{\Delta}  \frac{\partial \ln T_z}{\partial \theta},
\frac{\lambda_z}{qR}  \frac{L_{n_z}}{ \rho_{pz}} \frac{\partial \ln T_z}{\partial \theta} 
 \right\}
 \Delta \frac{ \lambda_z}{qR} \sqrt{\frac{z_z}{z_i}} \frac{\rho_{pz}}{L_{n_z}}
 \sim 
 \\
& \sim
 \frac{\partial \ln T_z}{\partial \theta}
\max \left\{
 \frac{\Delta}{\frac{\partial \ln T_z}{\partial \theta}} \frac{ \lambda_z}{qR} \sqrt{\frac{z_i}{z_z}} \frac{\rho_{pz}}{L_{n_z}}, 
   \frac{ \lambda_z}{qR} \sqrt{\frac{z_z}{z_i}} \frac{\rho_{pz}}{L_{n_z}},
 \Delta \left( \frac{\lambda_z}{qR} \right)^2  \sqrt{\frac{z_z}{z_i}}
 \right\}
 \ll 1.
 \label{eq:neglheatenergy}
 \end{aligned}
\end{equation}

Combining the most restrictive assumptions, equilibration will dominate when 
\begin{equation}
\Delta \frac{ \lambda_z}{qR} \sqrt{\frac{z_z}{z_i}}
\min \left\{
\frac{\rho_{pz}}{L_{n_z}},
 \frac{\lambda_z}{qR}
\right\}
\ll 1.
\end{equation} 
The impurity temperature is then equal to lowest order to the bulk ion temperature, which is taken to be a flux function.



%

\bibliography{main}

\end{document}